\documentclass[doublecol]{epl2}
\usepackage{amssymb,amsmath,graphicx,setspace,color,rotating,subfigure,url}
\usepackage{extarrows}
\title{Market correlation structure changes around the Great Crash}

\author{Rui-Qi Han\inst{1,2} \and Wen-Jie Xie\inst{2,3,4} \and Xiong Xiong\inst{5,6}\footnote{e-mail: xxpeter@tju.edu.cn} \and Wei Zhang\inst{5,6} \and Wei-Xing Zhou\inst{1,2,3}\footnote{e-mail: wxzhou@ecust.edu.cn}}
\shortauthor{R.-Q. Han \etal}

\institute{
  \inst{1} Department of Mathematics, East China University of Science and Technology, Shanghai 200237, China\\
  \inst{2} Research Center for Econophysics, East China University of Science and Technology, Shanghai 200237, China\\
  \inst{3} School of Business, East China University of Science and Technology, Shanghai 200237, China\\
  \inst{4} Postdoctoral Research Station, East China University of Science and Technology, Shanghai 200237, China\\
  \inst{5} College of Management and Economics, Tianjin University, Tianjin 300072, China\\
  \inst{6} China Center for Social Computing and Analytics, Tianjin University, Tianjin 300072, China
}


\abstract{
  We perform a comparative analysis of the Chinese stock market around the occurrence of the 2008 crisis based on the random matrix analysis of high-frequency stock returns of 1228 stocks listed on the Shanghai and Shenzhen stock exchanges. Both raw correlation matrix and partial correlation matrix with respect to the market index in two time periods of one year are investigated. We find that the Chinese stocks have stronger average correlation and partial correlation in 2008 than in 2007 and the average partial correlation is significantly weaker than the average correlation in each period. Accordingly, the largest eigenvalue of the correlation matrix is remarkably greater than that of the partial correlation matrix in each period. Moreover, each largest eigenvalue and its eigenvector reflect an evident market effect, while other deviating eigenvalues do not. We find no evidence that deviating eigenvalues contain industrial sectorial information. Surprisingly, the eigenvectors of the second largest eigenvalues in 2007 and of the third largest eigenvalues in 2008 are able to distinguish the stocks from the two exchanges. We also find that the component magnitudes of the some largest eigenvectors are proportional to the stocks' capitalizations.
}

\begin{document}

\maketitle


\section{Introduction}
\label{S1:Introduction}

Financial markets evolve in a self-organized manner with the interacting elements forming complex networks at different levels, including international markets \cite{Song-Tumminello-Zhou-Mantegna-2011-PRE,Kumar-Deo-2012-PRE,Nobi-Maeng-Ha-Lee-2013-JKPS,Nobi-Lee-Kim-Lee-2014-PLA}, individual markets \cite{Mantegna-1999-EPJB,Plerou-Gopikrishnan-Rosenow-Amaral-Stanley-1999-PRL,Laloux-Cizean-Bouchaud-Potters-1999-PRL,Kwapien-Drozdz-2012-PR}, and security trading networks \cite{Jiang-Zhou-2010-PA,Wang-Zhou-Guan-2011-PA,Sun-Cheng-Shen-Wang-2011-PA,Sun-Shen-Cheng-Wang-2012-PLoS1,Wang-Zhou-Guan-2012-Nc,Jiang-Xie-Xiong-Zhang-Zhang-Zhou-2013-QFL,Sun-Shen-Cheng-2014-SR,Li-Jiang-Xie-Xiong-Zhang-Zhou-2015-PA}. There are well-documented stylized facts of stock return time series within individual markets unveiled by the random matrix theory (RMT) analysis \cite{Plerou-Gopikrishnan-Rosenow-Amaral-Stanley-1999-PRL,Plerou-Gopikrishnan-Rosenow-Amaral-Guhr-Stanley-2002-PRE}: (1) The largest eigenvalue reflects the market effect such that its eigenportfolio returns are strongly correlated with the market returns; (2) Other largest eigenvalues contain information of industrial sectors; and (3) The smallest eigenvalues embed stock pairs with large correlations. However, for stock exchange index returns \cite{Song-Tumminello-Zhou-Mantegna-2011-PRE} and housing markets \cite{Meng-Xie-Jiang-Podobnik-Zhou-Stanley-2014-SR,Meng-Xie-Zhou-2015-IJMPB}, the largest eigenvalues can be used to extract geographic traits. Moreover, the signs of eigenvector components contain information of local interactions \cite{Jiang-Zheng-2012-EPL,Jiang-Chen-Zheng-2014-SR,Meng-Xie-Jiang-Podobnik-Zhou-Stanley-2014-SR}.

Financial crises occur more frequently than people usually expect, during which financial markets experience abrupt regime changes like phase transitions \cite{Sornette-2003,Sornette-2003-PR}. The market correlation structure changes around financial crashes. The average correlation after the critical point of crash is higher than that before the crash \cite{Plerou-Gopikrishnan-Rosenow-Amaral-Guhr-Stanley-2002-PRE,Nobi-Maeng-Ha-Lee-2013-JKPS,Ren-Zhou-2014-PLoS1}. Also, the correlation network becomes more connected after the crash \cite{Nobi-Maeng-Ha-Lee-2014-PA}. It is natural that the absorption ratio of the largest eigenvalue serves as a measure of systemic risk \cite{Billio-Getmansky-Lo-Pelizzon-2012-JFE}.

On the other hand, the partial correlation analysis, which is a powerful tools for investigating the intrinsic correlation between two time series effected by common factors \cite{Baba-Shibata-Sibuya-2004-ANZJS}, has been applied to financial markets \cite{Plerou-Gopikrishnan-Rosenow-Amaral-Guhr-Stanley-2002-PRE,Kenett-Shapira-BenJacob-2009-JPS,Kenett-Tumminello-Madi-GurGershgoren-Mantegna-BenJacob-2010-PLoS1,Meng-Xie-Jiang-Podobnik-Zhou-Stanley-2014-SR,Kenett-Huang-Vodenska-Havlin-Stanley-2015-QF,Uechi-Akutsu-Stanley-Marcus-Kenett-2015-PA,Qian-Liu-Jiang-Podobnik-Zhou-Stanley-2015-PRE,Meng-Xie-Zhou-2015-IJMPB}. An intriguing feature is that the partial correlation analysis is able to identify influences among different time series \cite{Kenett-Huang-Vodenska-Havlin-Stanley-2015-QF}.

Applying the random matrix theory analysis to the 1-min high-frequency returns of Chinese stocks, we investigate in this work the correlation structure changes around the breakout of the Great Crash at the end of 2007 \cite{Jiang-Zhou-Sornette-Woodard-Bastiaensen-Cauwels-2010-JEBO}. We focus on unveiling the information contents embedded in the deviating eigenvalues and their associated eigenvectors of the raw and partial correlation matrices. Novel results are found.

\section{Data sets}
\label{S1:Data}

We investigate the 1-min return time series of 489 A-share stocks traded on the Shenzhen Stock Exchange (SZSE) and 739 A-share stocks traded on the Shanghai Stock Exchange (SHSE) in 2007 and 2008, totally 1228 stocks, which were kindly provided by RESSET (http://resset.cn/). These stocks are chosen such that they have been listed before 2007 and had at least 180 trading days in each year.

The SZSE stocks belong to 16 industrial sectors including manufacturing (C, 306, 62.58\%), real estate (K, 39, 7.98\%), wholesale and retail industry (F, 39, 7.98\%), electric power, heat, gas and water production and supply (D, 26, 5.32\%), transport, storage and postal service (G, 14, 2.86\%), mining (B, 11, 2.25\%), information transmission, software and information technology services (I, 10, 2.04\%), and 9 other industries. The SHSE stocks belong to 18 industrial sectors including manufacturing (C, 397, 53.72\%), wholesale and retail industry (F, 74, 10.01\%), real estate (K, 54, 7.31\%), transport, storage and postal service (G, 47, 6.36\%), electric power, heat, gas and water production and supply (D, 41, 5.55\%), mining (B, 23, 3.11\%), information transmission, software and information technology services (I, 21, 2.84\%), and 11 other industries.

The 1-min logarithmic returns of stock $i$ are calculated as follows
\begin{equation}
  r_i(t) = \ln P_i(t)-\ln P_i(t-1),
  \label{Eq:Return}
\end{equation}
where $P_i(t)$ denotes the price of stock $i$ at time $t$ and $t=1,2,\cdots,T$. The returns are calculated at the intraday manner and no overnight returns are considered.

\section{Distributions of correlation coefficients and partial correlation coefficients}
\label{S1:PDF:Corr}

For each year, 2007 or 2008, we calculate the correlation coefficient $c_{ij}$ between the returns of stock $i$ and stock $j$, which form the correlation matrix $\mathbf{C}$, as follows:
\begin{equation}
  c_{ij} = \frac{\langle{[r_i - \langle{r_i}\rangle][r_j - \langle{r_j}\rangle]\rangle}}{\sigma_{r_i}\sigma_{r_j}},
  \label{Eq:Cij}
\end{equation}
where $\sigma_{r_i}$ and $\sigma_{r_j}$ are the standard deviations of $r_i(t)$ and $r_j(t)$. It is common that the 1-min trading time sequences of stock $i$ and stock $j$ do not overlap. Under such circumstance, we discard those times appeared in only one stock. For two arbitrary return time series $r_i(t)$ and $r_j(t)$, we can extract their idiosyncratic components $\varepsilon_i(t)$ by removing a common collective component $r_m(t)$ and calibrating the following simple univariate factor model:
\begin{equation}
  r_i(t) = \alpha_i + \beta_ir_m(t) + \varepsilon_i(t),
\end{equation}
where $r_m(t)$ is the 1-min return time series of the Shanghai Stock Exchange Composite Index (SSCI). The partial correlation coefficient $\rho_{ij}$ between $r_i(t)$ and $r_j(t)$ with respect to $r_m(t)$ is defined as the correlation coefficient between the two residuals $\varepsilon_i(t)$ and $\varepsilon_j(t)$:
\begin{equation}
  \rho_{ij} = \frac{\langle{[\varepsilon_i - \langle\varepsilon_i\rangle][\varepsilon_j - \langle\varepsilon_j\rangle]}\rangle}{\sigma_{\varepsilon_i}\sigma_{\varepsilon_j}},
  \label{Eq:Pij}
\end{equation}
where $\sigma_{\varepsilon_i}$ and $\sigma_{\varepsilon_j}$ are the standard deviations of $\varepsilon_i(t)$ and $\varepsilon_j(t)$. We denote $\mathbf{P}=[\rho_{ij}]$ the partial correlation matrix. Simple algebraic manipulations result in the following equation
\cite{Baba-Shibata-Sibuya-2004-ANZJS,Kenett-Tumminello-Madi-GurGershgoren-Mantegna-BenJacob-2010-PLoS1}
\begin{equation}
  \rho_{ij} = \frac{c_{ij}-c_{im}c_{jm}} {\sqrt{\left(1-c_{im}^2\right)\left(1-c_{jm}^2\right)}},
\end{equation}
where $c_{im}$ ($c_{jm}$) is the correlation coefficient between $r_i$ ($r_j$) and $r_m$.

\begin{figure}[ht]
  \centering
  \includegraphics[width=0.9\linewidth]{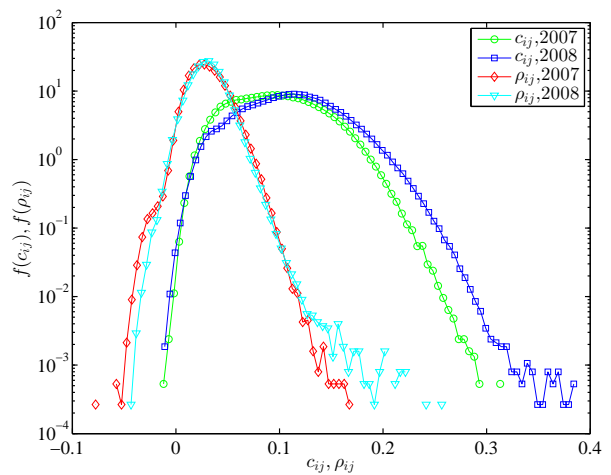}
  \caption{(Color online.) Probability distributions $f(c_{ij})$ and $f(\rho_{ij})$ of correlation coefficients $c_{ij}$ and partial correlation coefficients $\rho_{ij}$ of stock return time series in 2007 and 2008.}\label{Fig:RMT:Ret:1min:PDF:Cij}
\end{figure}

Fig.~\ref{Fig:RMT:Ret:1min:PDF:Cij} illustrates that the distributions $f(c_{ij})$ and $f(\rho_{ij})$ of correlation coefficients $c_{ij}$ and partial correlation coefficients $\rho_{ij}$ in 2007 and 2008. The overwhelming majority of $c_{ij}$ are positive and the distributions are rightly skewed. The average correlation coefficient in 2007 ($\langle{c_{ij,2007}}\rangle=0.098$) is slightly smaller than that in 2008 ($\langle{c_{ij,2008}}\rangle=0.112$), implying that the Chinese stock market was in high risk and this systemic risk is higher in the panic bearish period than in the mania bullish period. The average correlation coefficients of 1-min returns are significantly smaller than those of daily returns (about 0.35) \cite{Shen-Zheng-2009a-EPL,Ren-Zhou-2014-PLoS1}, which is due to the fact that high-frequency returns are more noisy. The maximum correlation coefficient is 0.314 in 2007 and 0.386 in 2008.

After removing the influence of SSCI, the distributions $f(\rho_{ij,2007})$ and $f(\rho_{ij,2008})$ become narrower with the average partial correlation coefficients closer to 0 when compared with $f(c_{ij,2007})$ and $f(c_{ij,2008})$, the bulk parts with $-0.02<\rho<0.12$ of the two distributions almost overlap, and the discrepancy between the two distributions $f(\rho_{ij,2007})$ and $f(\rho_{ij,2008})$ becomes much smaller. It suggests that the discrepancy between $f(c_{ij,2007})$ and $f(c_{ij,2008})$ is mainly caused by a market effect. We further observe that there are more negative partial correlation coefficients in 2007 and more positive partial correlation coefficients in 2008. However, the proportions are low.

\section{Distributions of eigenvalues}
\label{S1:PDF:Eigenvalues}

For the correlation matrix and the partial correlation matrix, we can determine their eigenvalues and the associated eigenvectors. When the observed time series have zero mean and unit variance, in the limit $N\rightarrow\infty$, $T\rightarrow\infty$ where $Q=T/N\geq1$ is fixed, the random matrix theory predicts that the distribution $f_{\rm{rmt}}(\lambda)$ of eigenvalues $\lambda$ can be expressed as \cite{Edelman-1988-SIAMjmaa,Sengupa-Mitra-1999-PRE}
\begin{equation}
  f_{\rm{{rmt}}}(\lambda) = \frac{Q}{2\pi}\frac{\sqrt{(\lambda_{\max}-\lambda)(\lambda-\lambda_{\min})}}{\lambda}
  \label{Eq:RMT:eigenvalue:density}
\end{equation}
for $\lambda\in[\lambda_{\min},\lambda_{\max}]$, where $\lambda_{\min}$ and $\lambda_{\max}$ are respectively given by
\begin{equation}
  \lambda^{\max}_{\min} = \left(1 \pm \sqrt{1/Q}\right)^2,
  \label{Eq:RMT:eigenvalue:range}
\end{equation}
which predicts a finite range of eigenvalues determined by the ratio $Q=T/N$.

There are $N=1228$ stocks in our sample. For the raw and partial correlation matrices in 2007, $T=56882$ and thus $Q=46.32$. We obtain that $\lambda_{\min}=0.728$ and $\lambda_{\max}=1.315$. For the raw and partial correlation matrices in 2008, $T=58310$ and thus $Q=47.48$. We obtain that $\lambda_{\min}=0.731$ and $\lambda_{\max}=1.311$. These characteristic values are summarized in Table \ref{TB:lambda}. Note that the results are essentially the same if we construct the random matrix from shuffled return time series \cite{Zhou-Mu-Kertesz-2012-NJP,Meng-Xie-Jiang-Podobnik-Zhou-Stanley-2014-SR}.

\begin{table}[tb]
  \centering
  \caption{Characteristic values of the eigenvalue distributions of the raw and partial correlation matrices in 2007 and 2008. $p_{<\lambda_{\min}}$ and $p_{>\lambda_{\max}}$ are respectively the percentages of empirical eigenvalues which are less than $\lambda_{\min}$ and greater than $\lambda_{\max}$.}
  \label{TB:lambda}
  \medskip
  \begin{tabular}{cccccccccccccccccccccc}
  \hline
                  & $\lambda_{\min}$ & $\lambda_{\max}$ & $\lambda_1$ & $p_{<\lambda_{\min}}$ & $p_{>\lambda_{\max}}$ \\\hline
  $c_{2007}$      & 0.728 & 1.315 & 130.8 & 27.04\% & 3.75\% \\
  $c_{2008}$      & 0.731 & 1.311 & 148.9 & 28.34\% & 2.20\% \\
  $\rho_{2007}$   & 0.728 & 1.315 & 41.89 & 15.39\% & 7.00\% \\
  $\rho_{2008}$   & 0.731 & 1.311 & 41.90 & 14.01\% & 5.94\% \\
  \hline
  \end{tabular}
\end{table}

\begin{figure}[bht]
  \centering
  \includegraphics[width=0.9\linewidth]{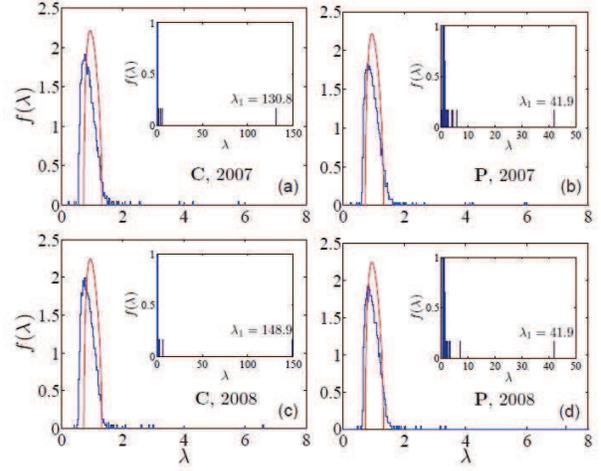}
  \caption{(Color online.) Probability distributions $f(\lambda)$ of the eigenvalues obtained from the raw and the partial correlation matrices of stock return time series in 2007 and 2008. The red smooth curve in each plot is the eigenvalue distribution predicted by the random matrix theory. The insets show the largest eigenvalues $\lambda_{1}$.}
  \label{Fig:RMT:Ret:1min:PDF:lambda}
\end{figure}

To identify that the estimated cross-correlations between stock returns are not a result of randomness, we compare in Fig.~\ref{Fig:RMT:Ret:1min:PDF:lambda} the empirical eigenvalue distributions $f(\lambda)$ of the raw and partial correlation matrices $\mathbf{C}$ and  $\mathbf{P}$ with the RMT prediction $f_{\mathrm{rmt}}(\lambda)$ given by Eq.~(\ref{Eq:RMT:eigenvalue:density}). It is somewhat ``trivial'' to observe that all the empirical eigenvalue distributions deviate significantly from the RMT prediction.

For the raw correlation matrices $\bf{C}$, there are 46 eigenvalues (3.75\%) exceeding $\lambda_{\max}=1.315$ in 2007 and 27 eigenvalues (2.20\%) exceeding $\lambda_{\max}=1.311$ in 2008. However, the largest eigenvalue $\lambda_1=148.9$ in 2008 is greater than $\lambda_1=130.8$ in 2007. Since $\lambda_1/N$ is a measure of systemic risk \cite{Billio-Getmansky-Lo-Pelizzon-2012-JFE}, we argue that the Chinese stock market has a higher systemic risk in 2008 than in 2007. Moreover, the largest eigenvalue captures 11.6\% of the variations in the return time series in 2007 and 12.1\% of the variations in 2008. We also find that 332 eigenvalues (27.04\%) are less than $\lambda_{\min}=0.728$ in 2007 and 348 eigenvalues (28.34\%) are less than $\lambda_{\min}=0.731$ in 2008. All these deviating eigenvalues and the associated eigenvectors might contain significant economic information.

For the partial correlation matrices $\bf{P}$, there are 86 eigenvalues (7.00\%) exceeding $\lambda_{\max}=1.315$ in 2007 and 73 eigenvalues (5.94\%) exceeding $\lambda_{\max}=1.311$ in 2008. Surprisingly, the largest eigenvalue $\lambda_1=41.90$ in 2008 is almost equal to $\lambda_1=41.89$ in 2007, indicating that the higher systemic risk in 2008 was mainly introduced by the co-movements of stocks. In addition, the largest eigenvalue accounts for about 3.4\% of the variations in the return residual time series in 2007 and 2008. We also find that 189 eigenvalues (15.39\%) are less than $\lambda_{\min}=0.728$ in 2007 and 172 eigenvalues (14.01\%) are less than $\lambda_{\min}=0.731$ in 2008. Overall, after removing the influence of SSCI, the eigenvalue distribution becomes much closer to the RMT prediction.

\begin{figure*}[htb]
  \centering
  \includegraphics[width=0.19\linewidth]{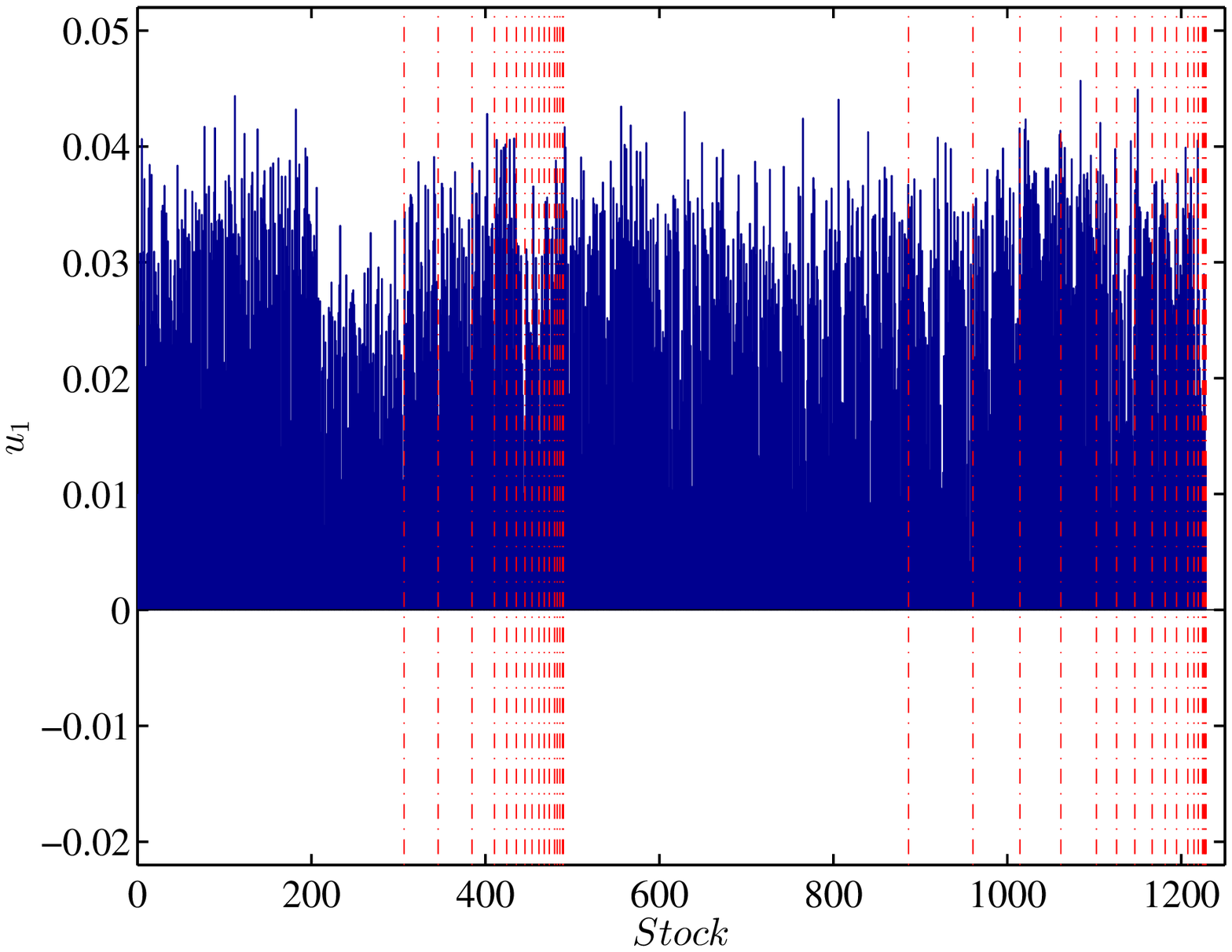}
  \includegraphics[width=0.19\linewidth]{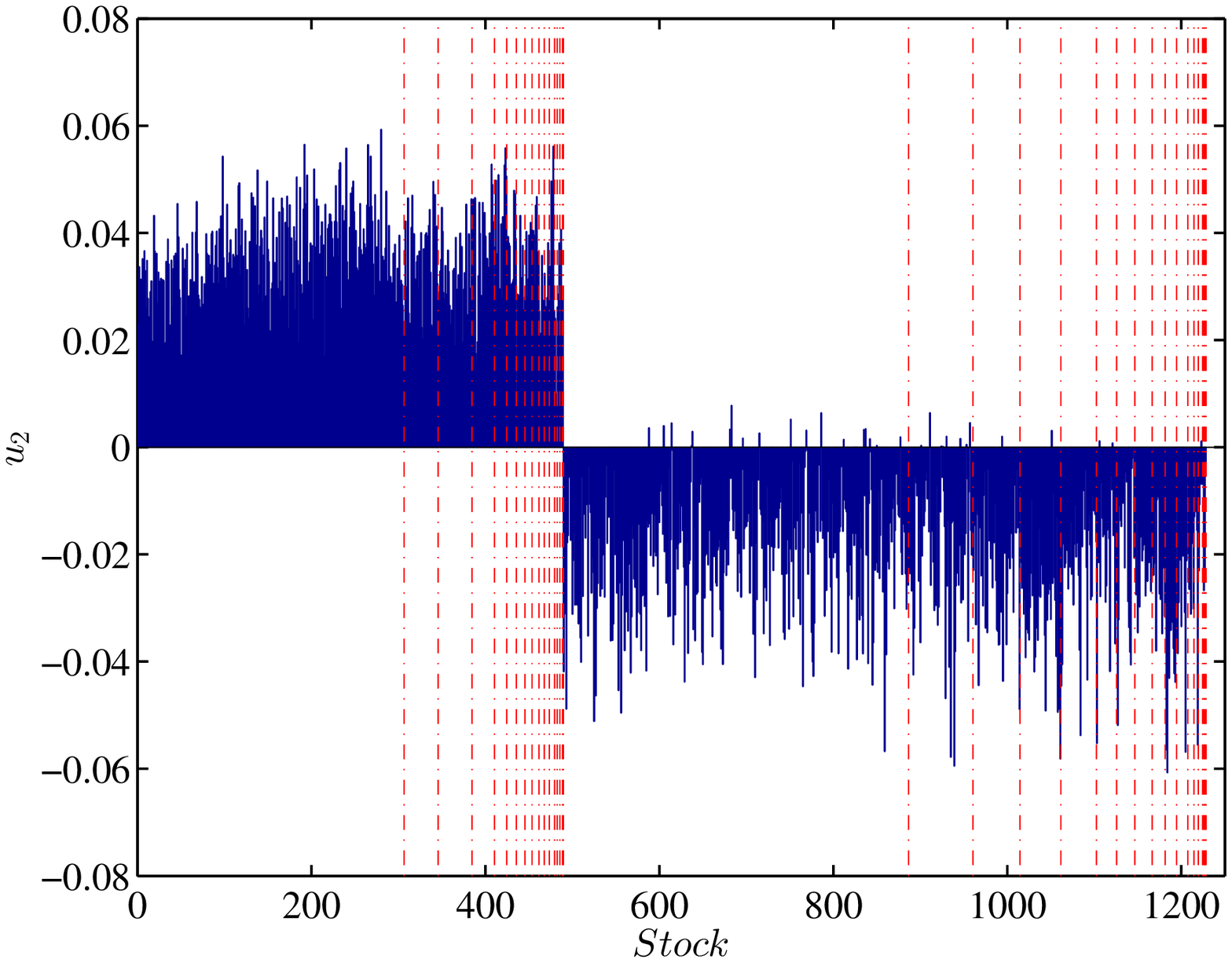}
  \includegraphics[width=0.19\linewidth]{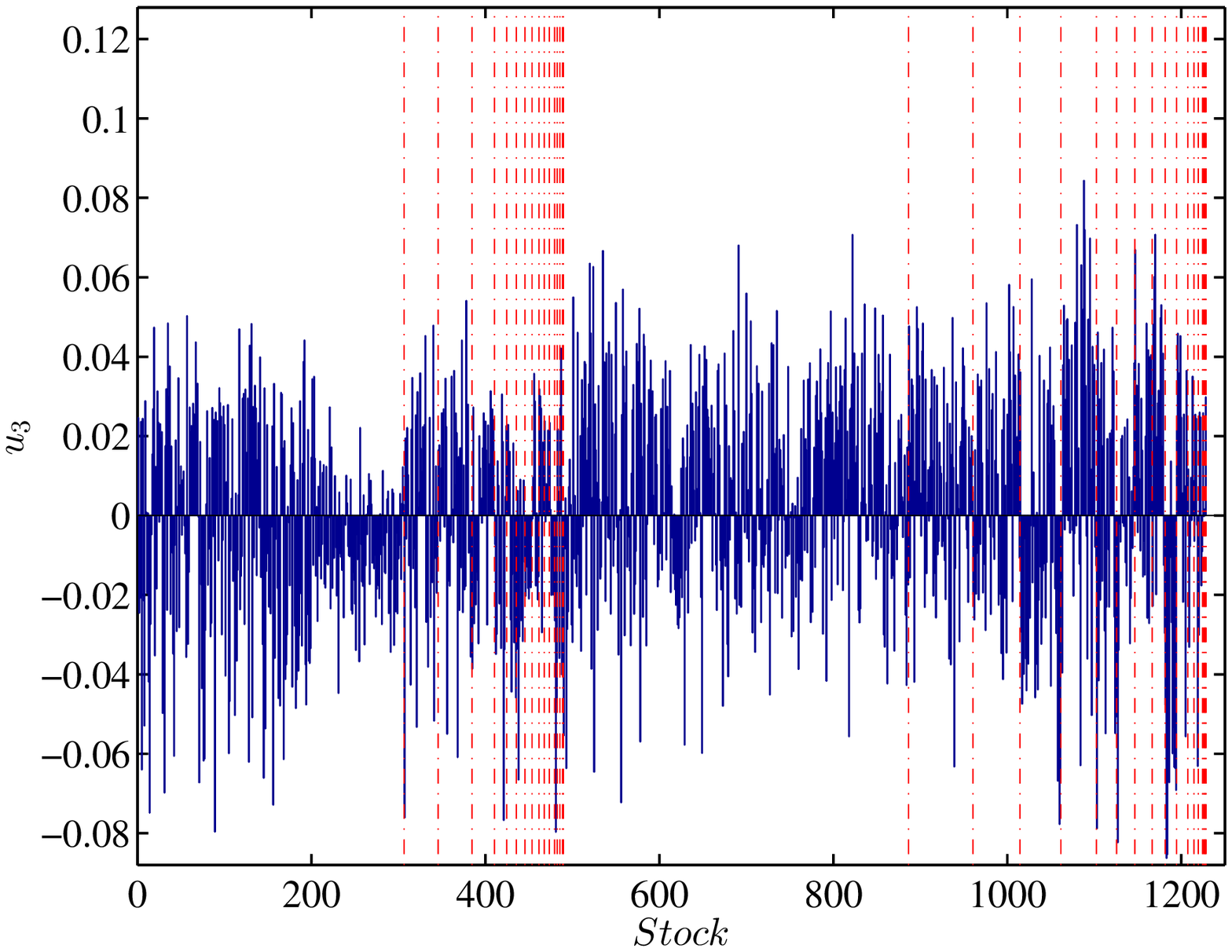}
  \includegraphics[width=0.19\linewidth]{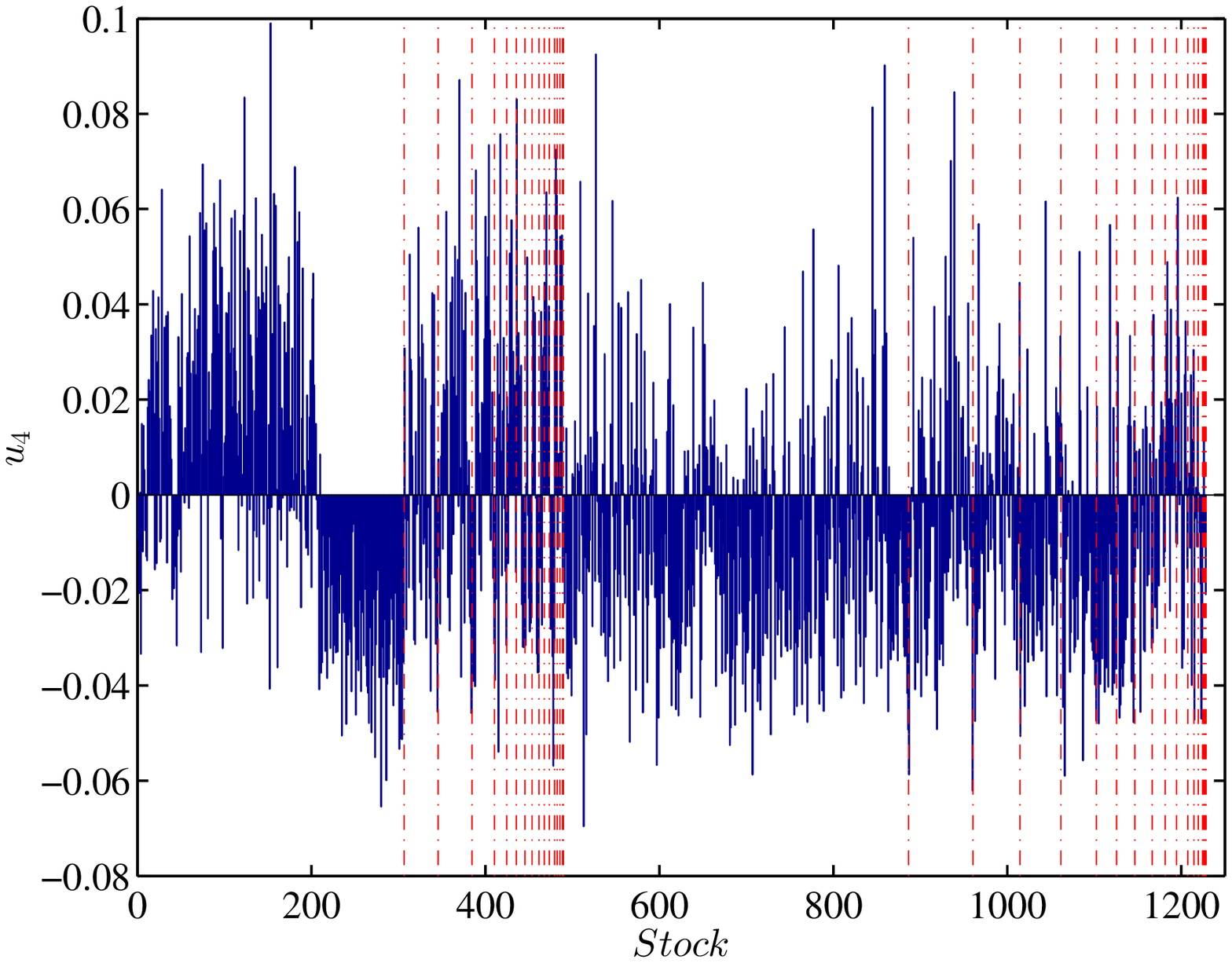}
  \includegraphics[width=0.19\linewidth]{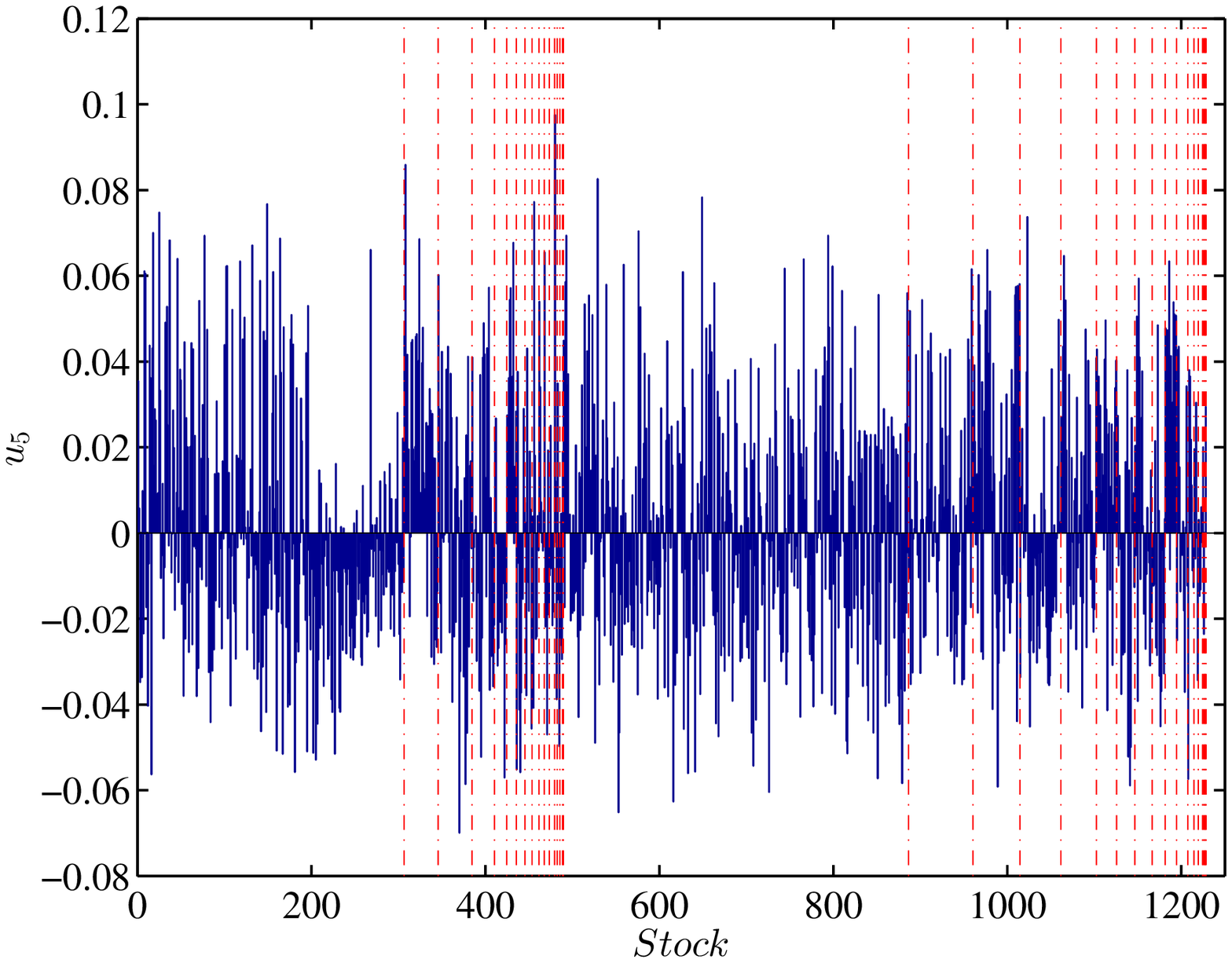}
  \includegraphics[width=0.19\linewidth]{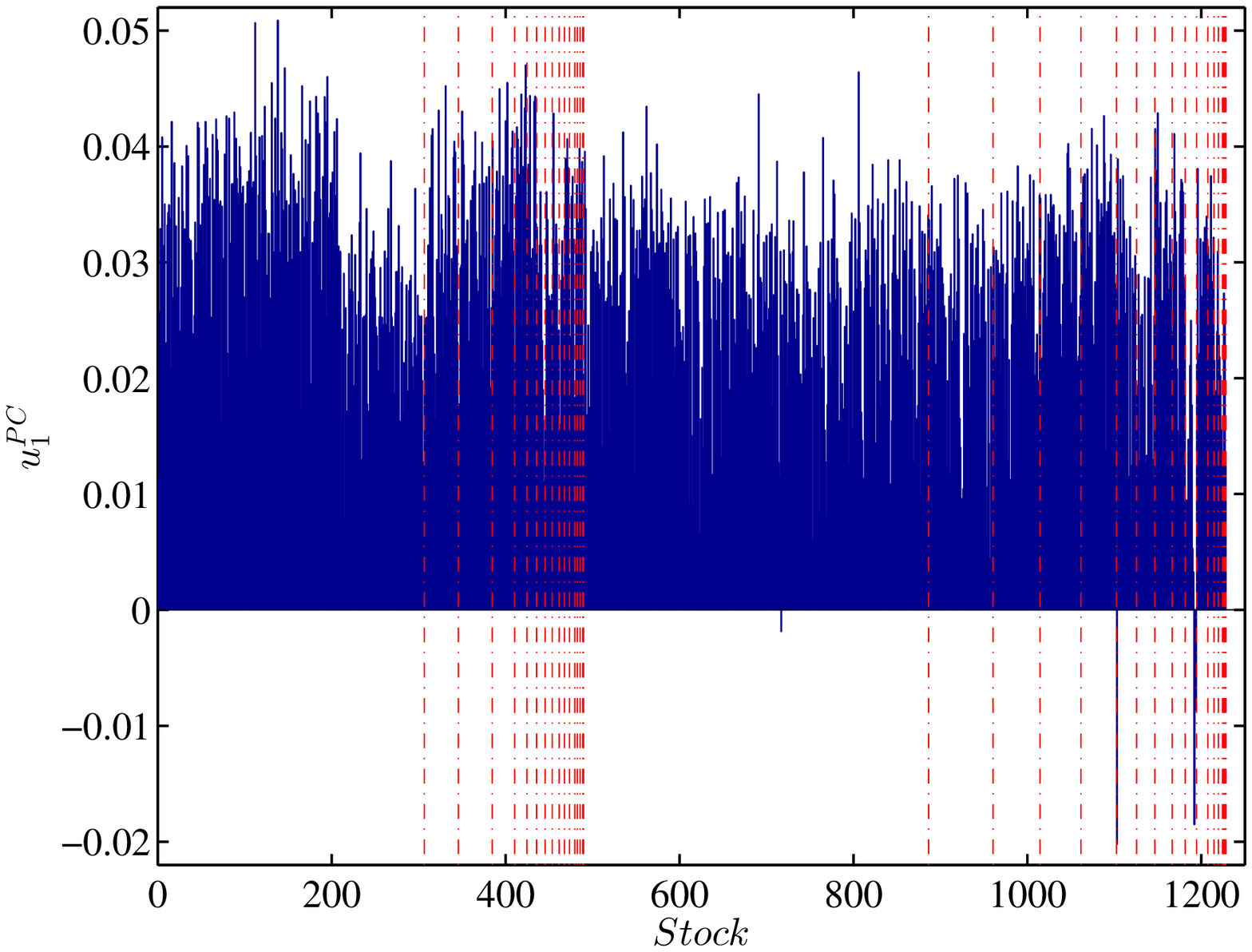}
  \includegraphics[width=0.19\linewidth]{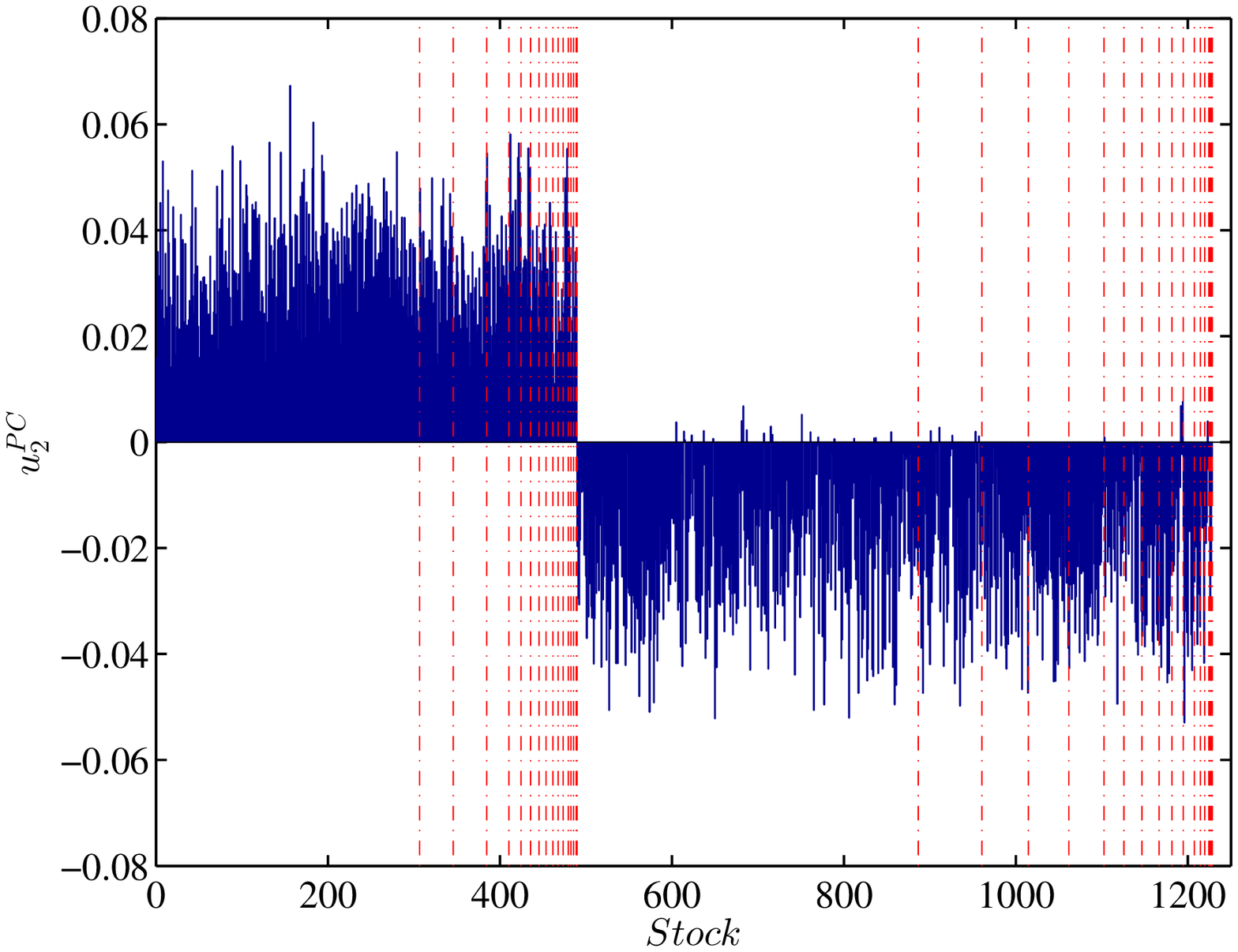}
  \includegraphics[width=0.19\linewidth]{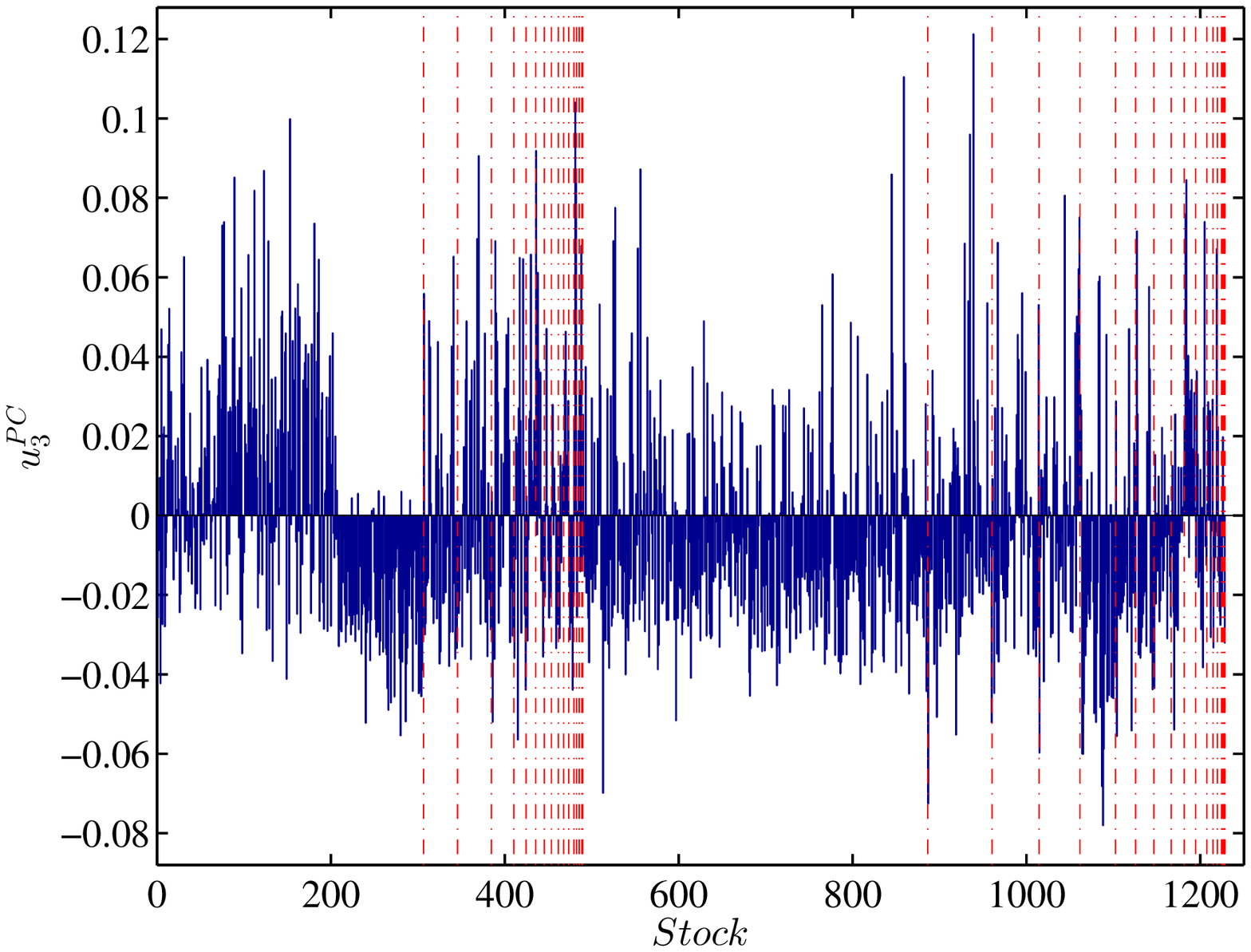}
  \includegraphics[width=0.19\linewidth]{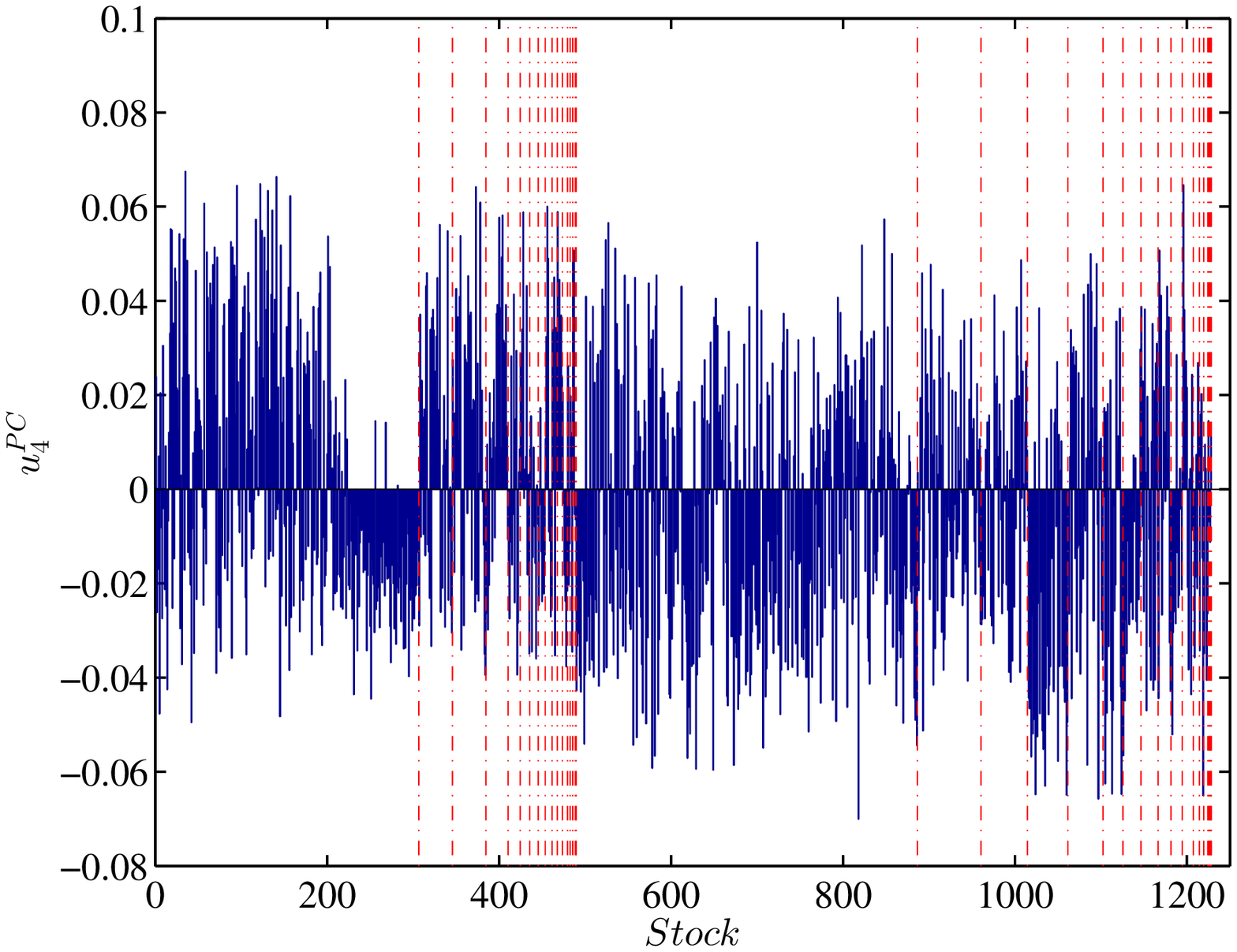}
  \includegraphics[width=0.19\linewidth]{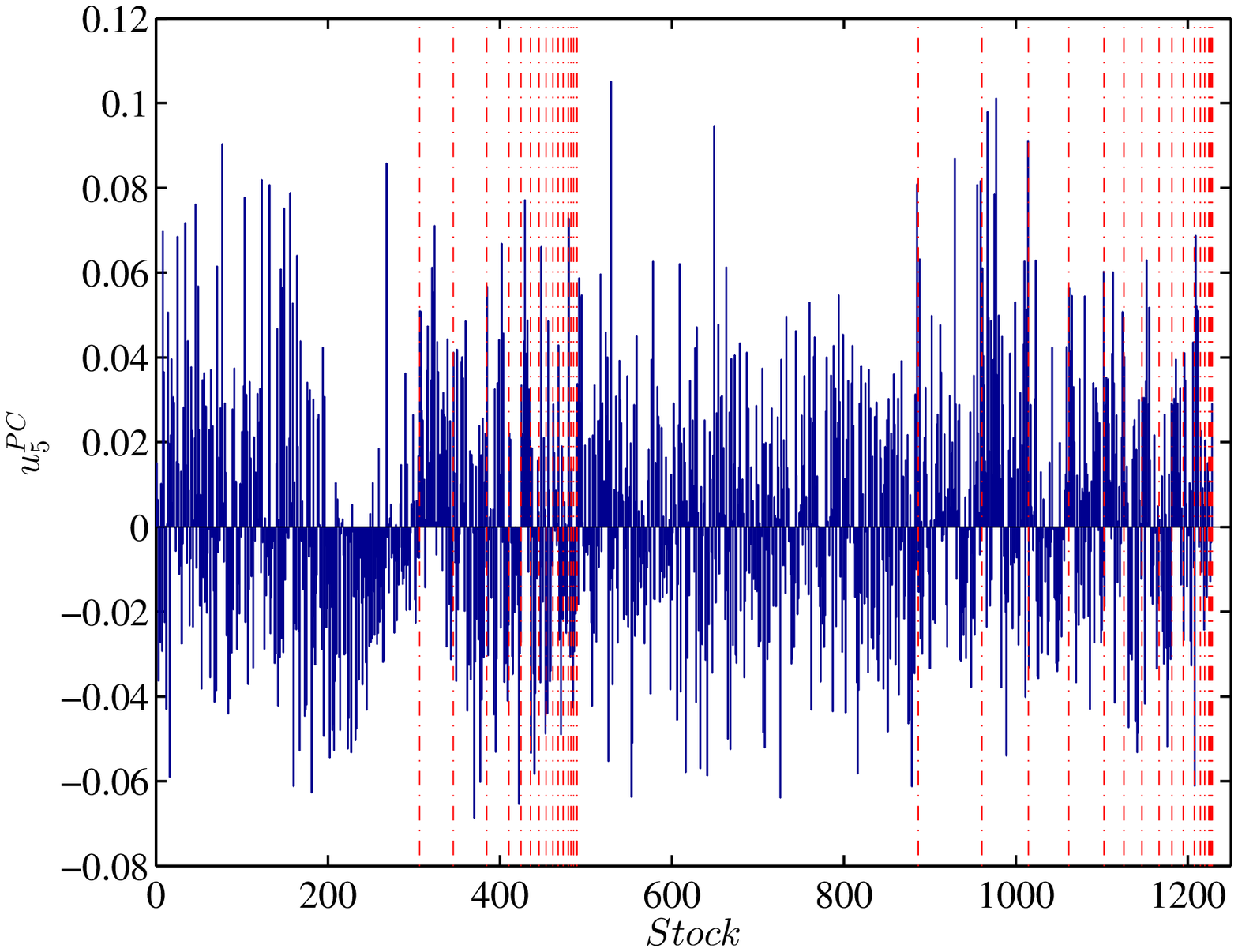}
  \includegraphics[width=0.19\linewidth]{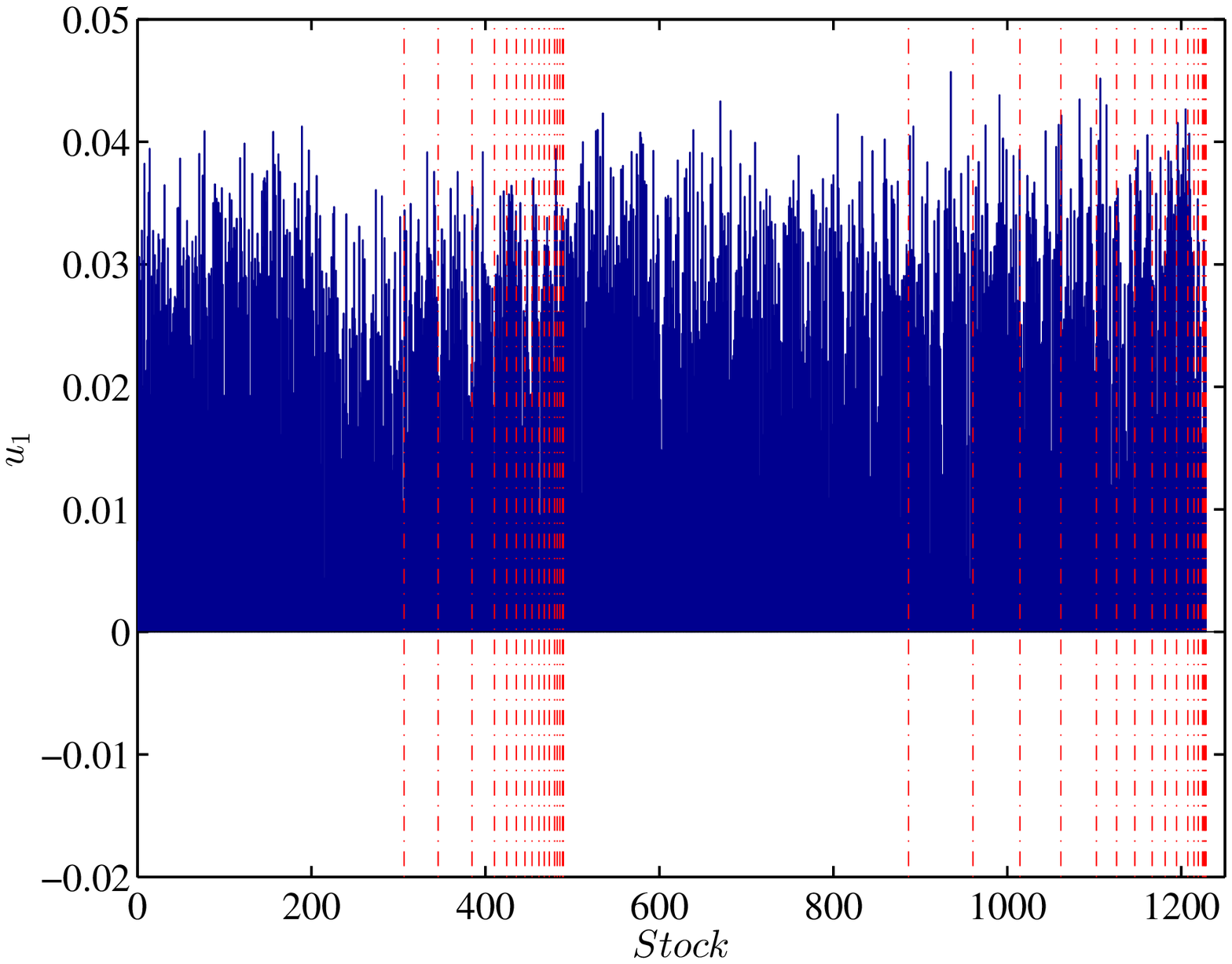}
  \includegraphics[width=0.19\linewidth]{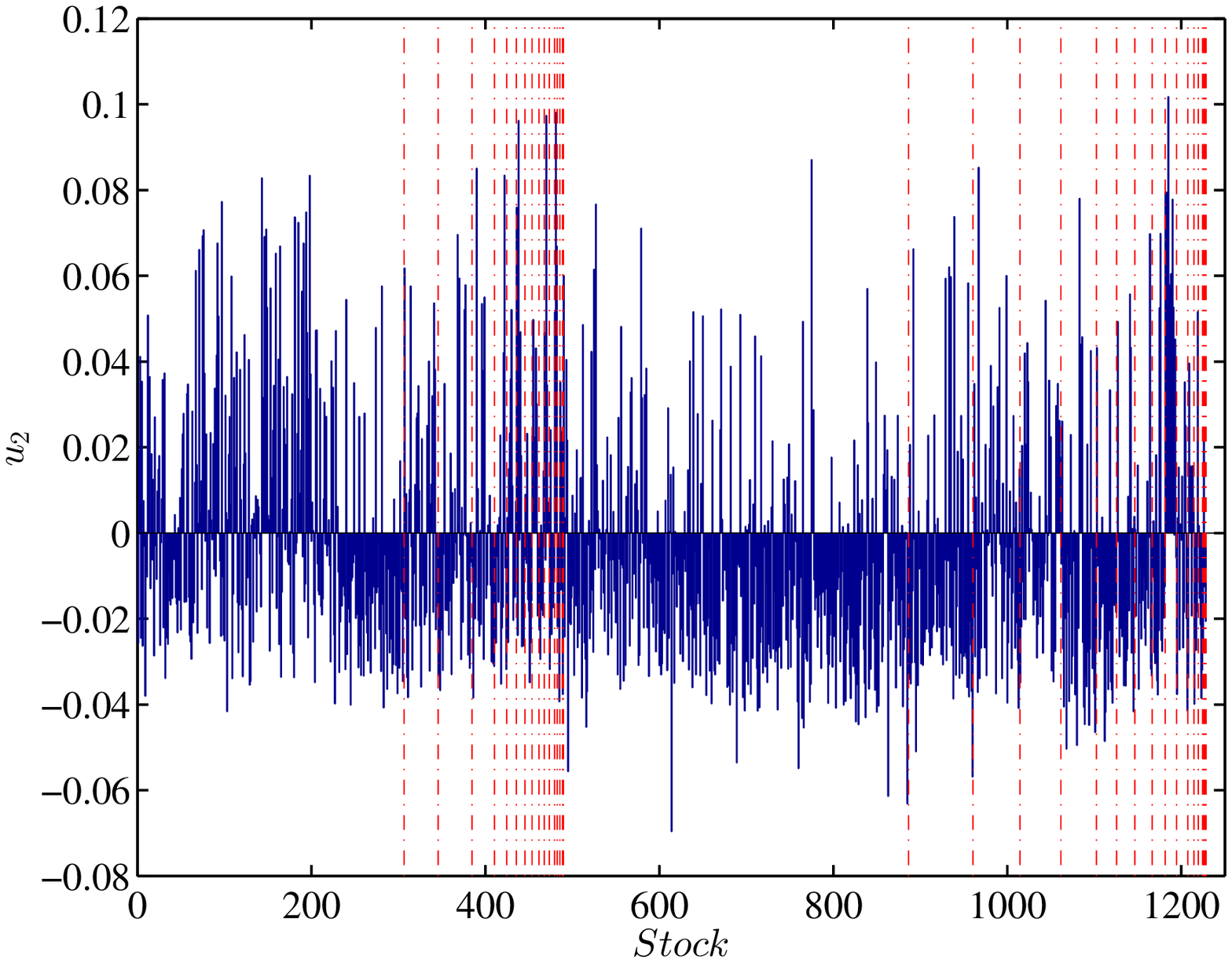}
  \includegraphics[width=0.19\linewidth]{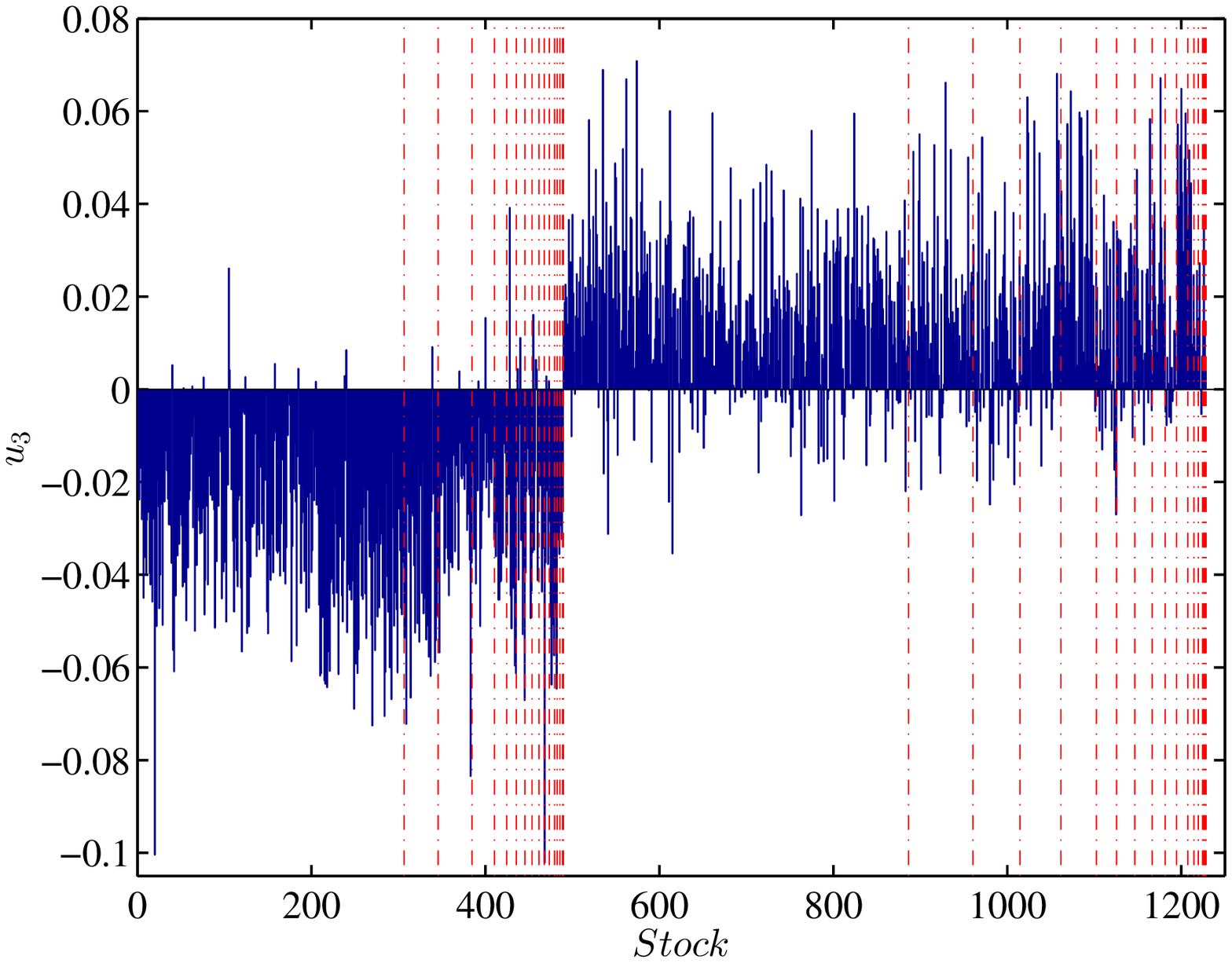}
  \includegraphics[width=0.19\linewidth]{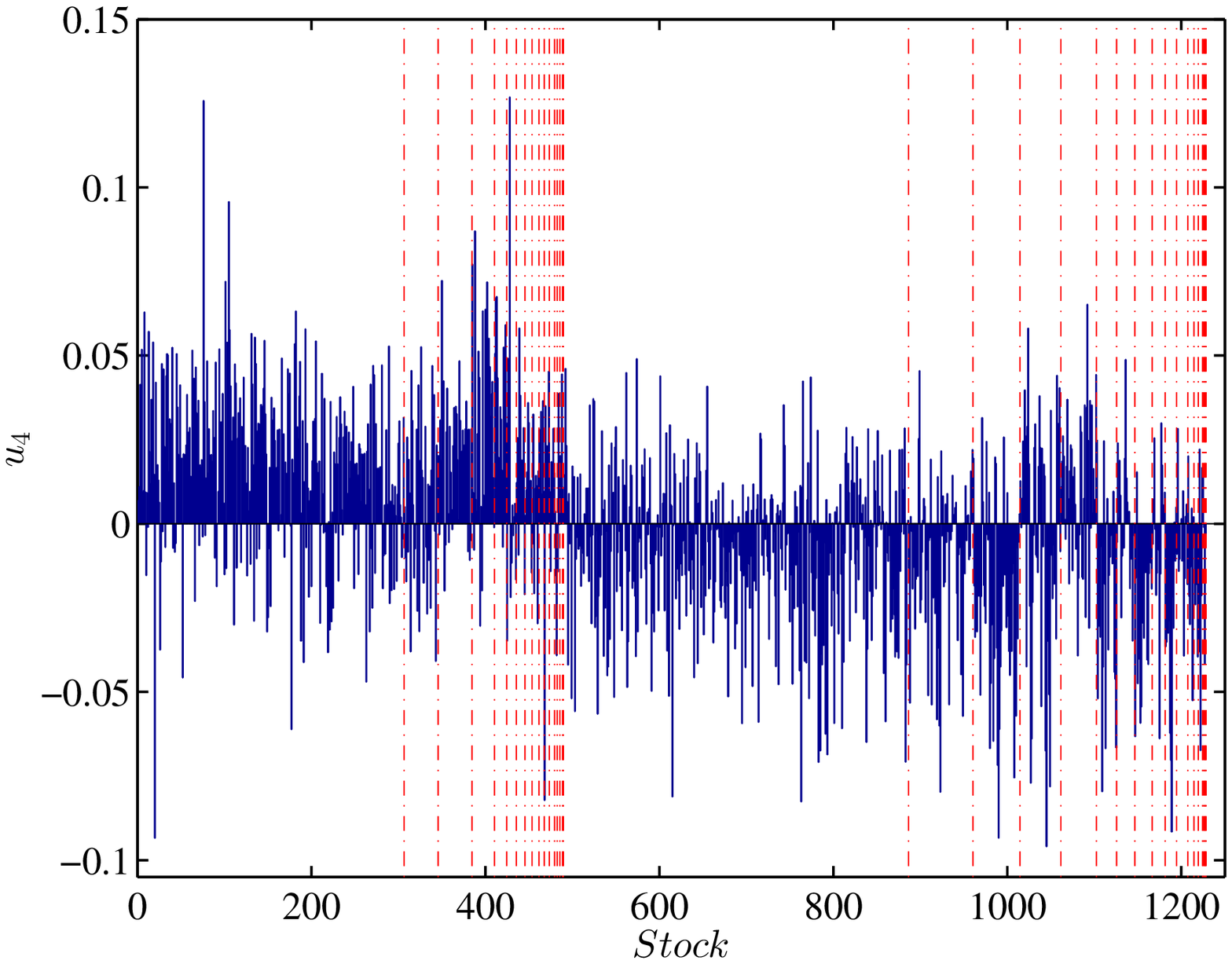}
  \includegraphics[width=0.19\linewidth]{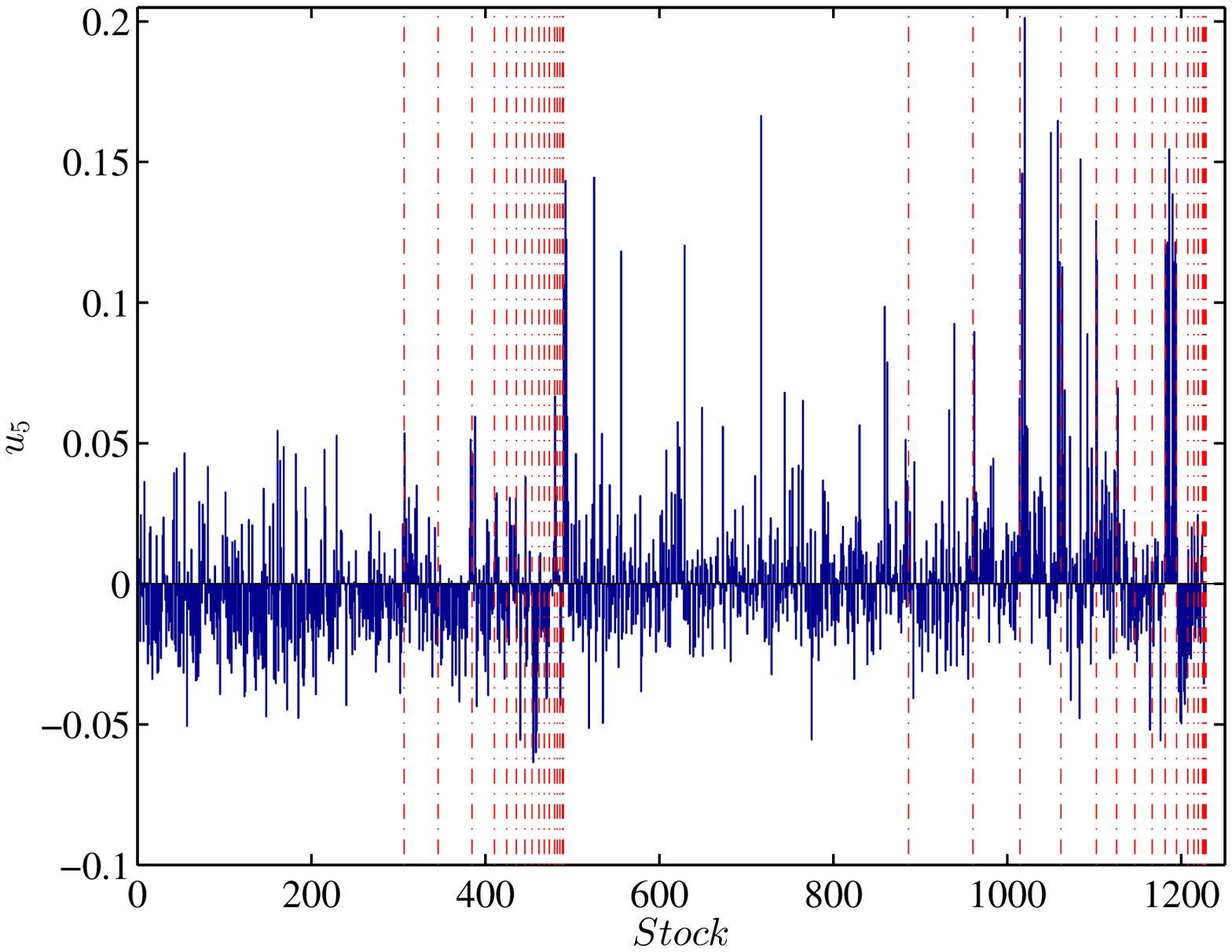}
  \includegraphics[width=0.19\linewidth]{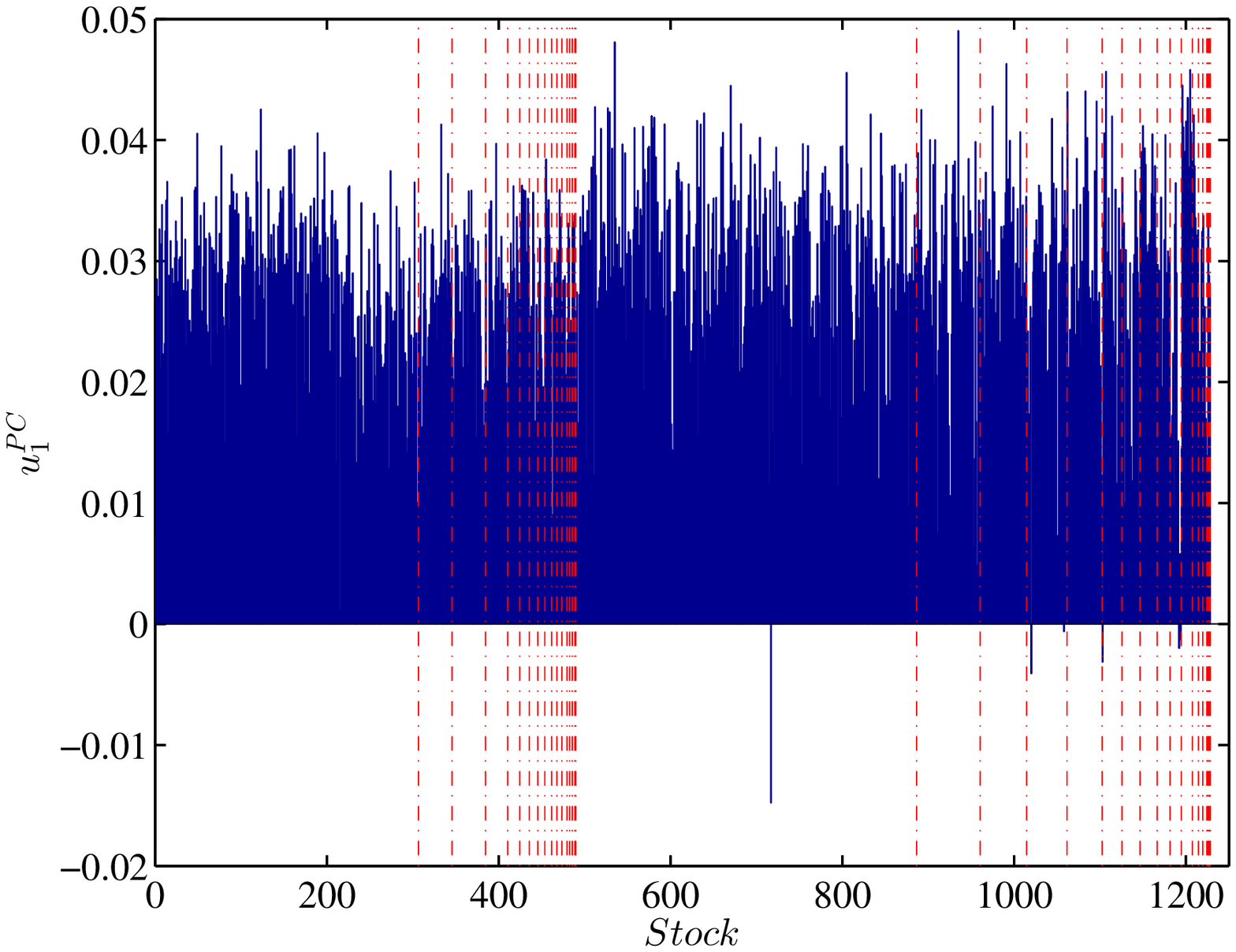}
  \includegraphics[width=0.19\linewidth]{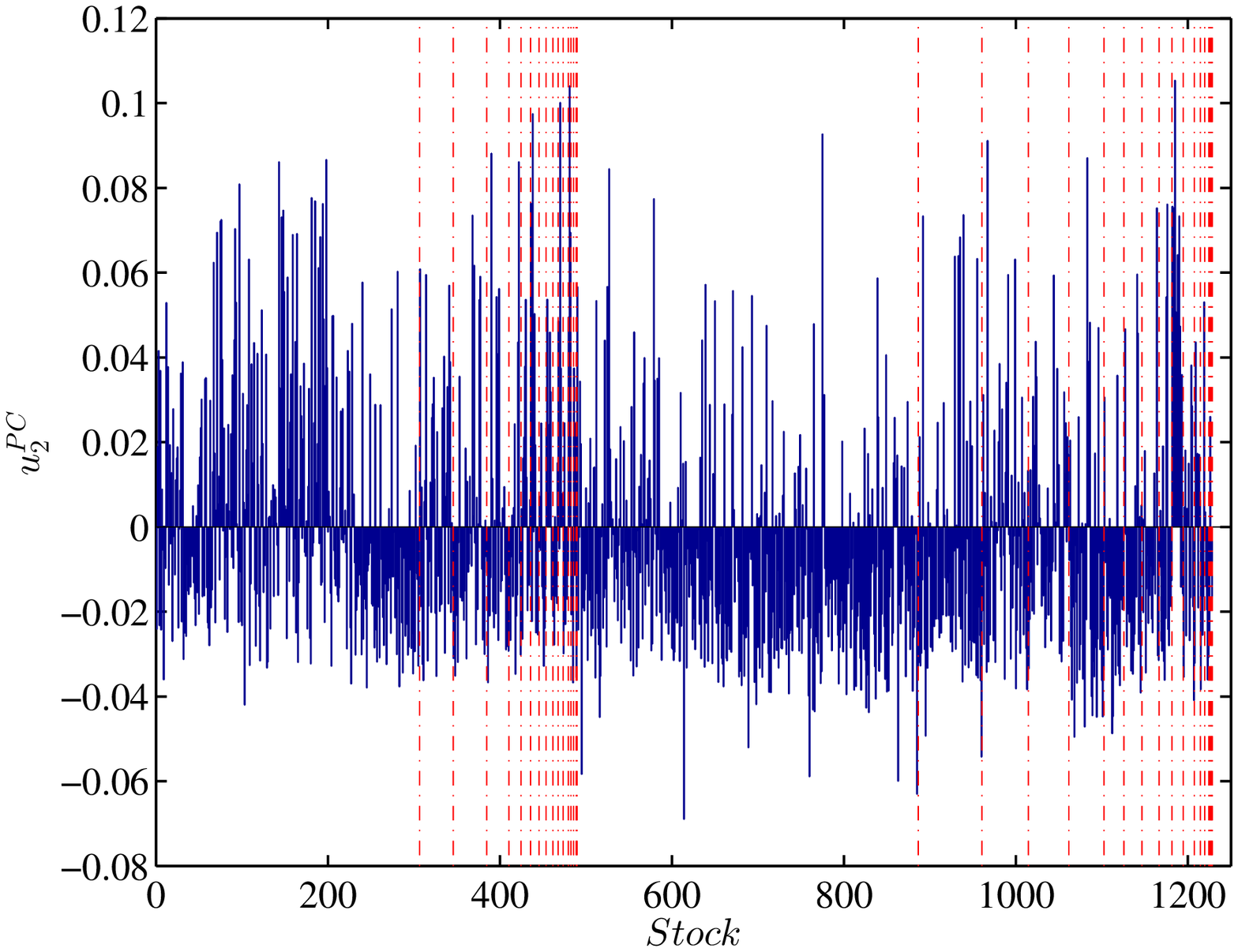}
  \includegraphics[width=0.19\linewidth]{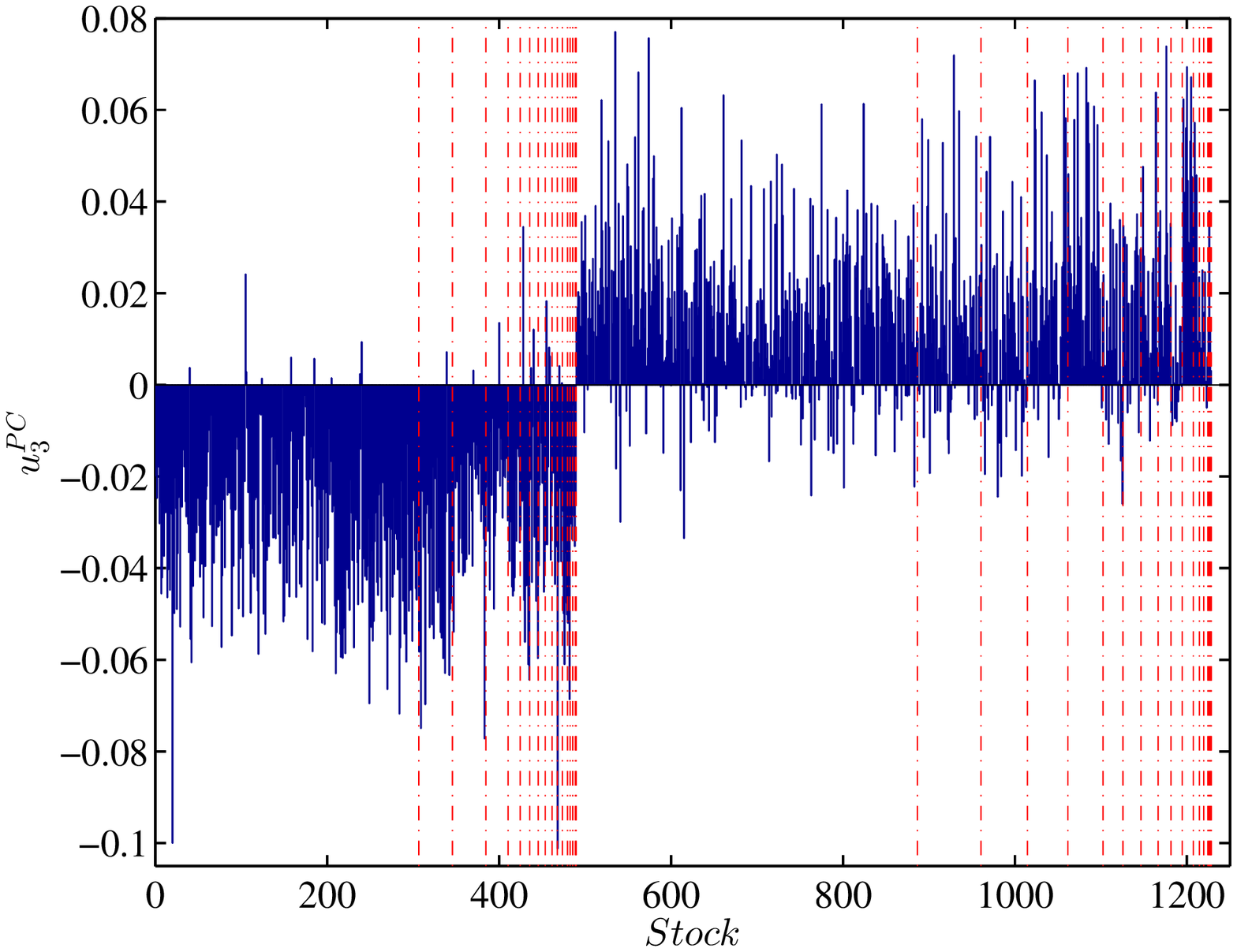}
  \includegraphics[width=0.19\linewidth]{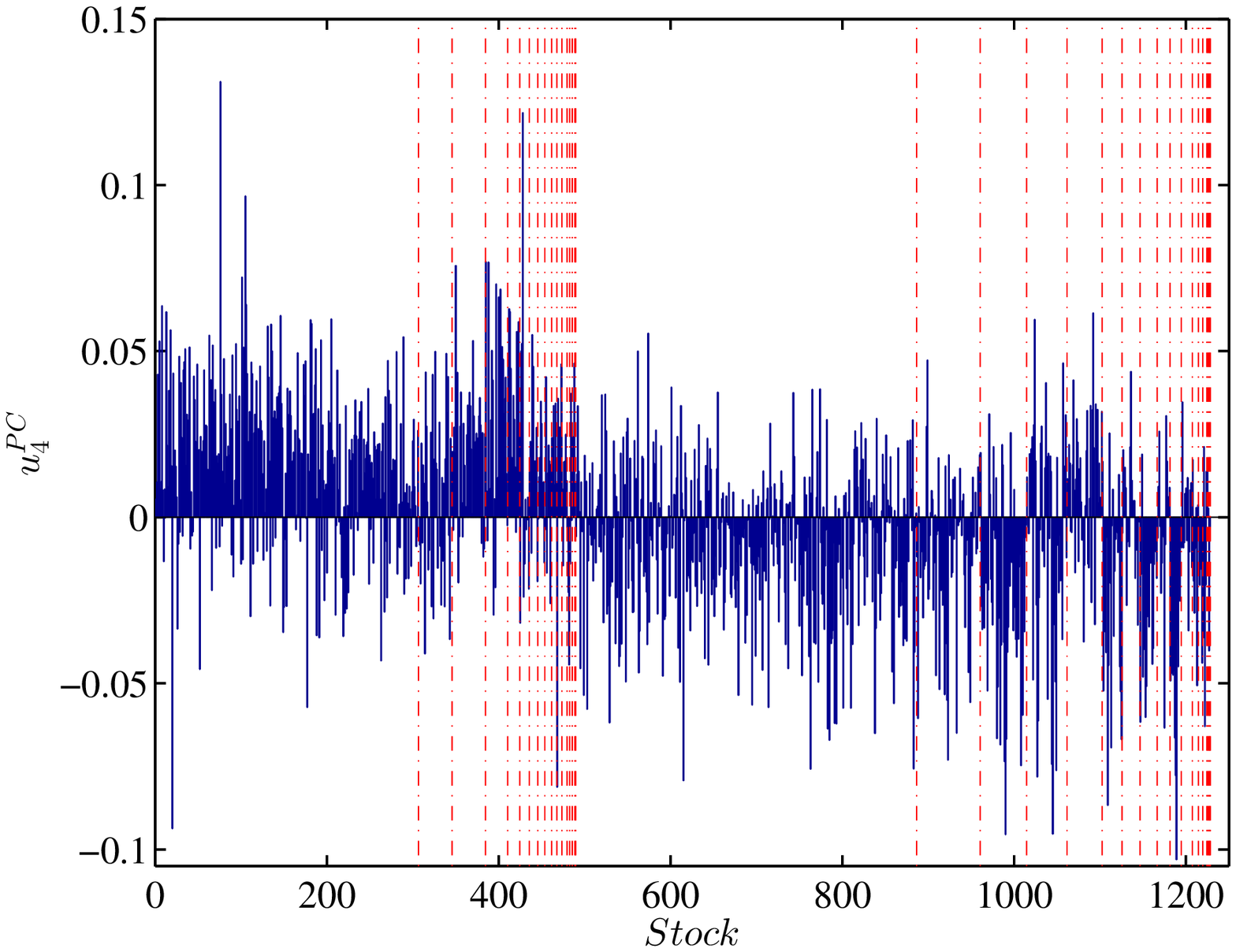}
  \includegraphics[width=0.19\linewidth]{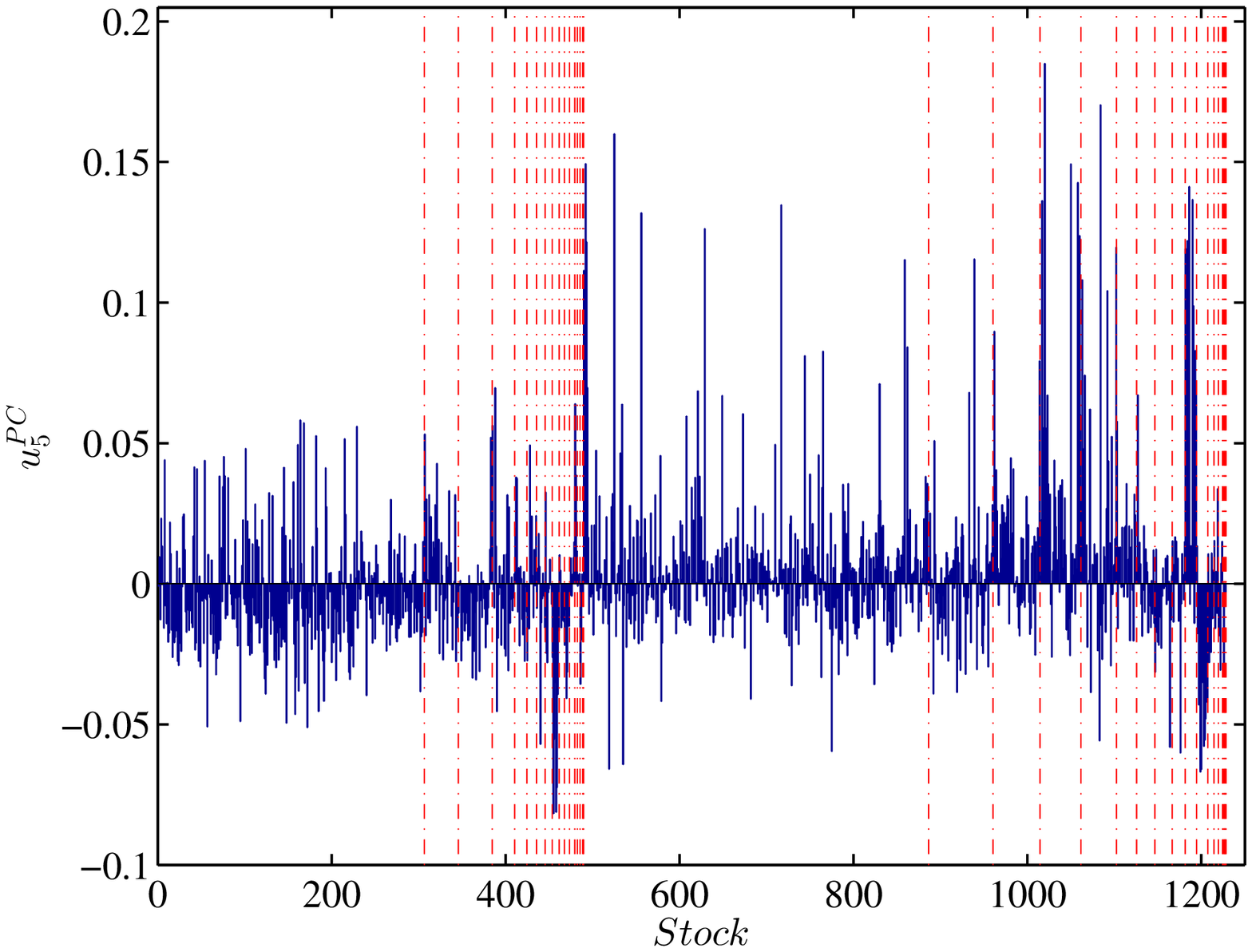}
  \caption{Components of the eigenvectors $u_1$ to $u_5$ corresponding to the five largest eigenvalues of the raw and partial correlation matrices in 2007 and 2008. The vertical dashed lines are used to separate in turn 489 SZSE stocks (39.8\%) and 739 SHSE stocks (60.2\%) according to their industries.}
  \label{Fig:RMT:Ret:1min:u1:u5}
\end{figure*}

\section{Eigenvectors of the largest eigenvalues}

We now turn to unveil the economic information embedded in the first few largest eigenvalues of each matrix. Fig.~\ref{Fig:RMT:Ret:1min:u1:u5} shows the associated eigenvectors of the five largest eigenvalues of the raw and partial correlation matrices in 2007 and 2008. Strikingly, there are no significant differences between the two eigenvectors associated with the $k$-th largest eigenvalue of the raw and partial correlation matrices in a given year. However, we observe significant differences for the eigenvectors in different years. We thus focus on the discussions of the raw correlation matrices below.

\begin{table}[htb]
  \centering
  \caption{Percentages of stocks traded on the SZSE and the SHSE whose components are positive (or negative) in the eigenvector $u_2$ or $u_3$ of the raw correlation matrix $\bf{C}$ in 2007 or 2008.}
  \label{TB:u2u3}
  \medskip
  \begin{tabular}{cccccccccccccccccccccc}
  \hline
   & & \multicolumn{2}{c}{2007} && \multicolumn{2}{c}{2008} \\
   \cline{3-4} \cline{6-7}
      &      & $+$ & $-$ && $+$ & $-$ \\\hline
$u_2$ & SZSE & 93.86\% & 0 && 49.64\% & 34.77\%  \\
      & SHSE &  6.14\% & 100\% &&  50.36\% & 65.23\%  \\
$u_3$ & SZSE &  35\% & 45\% &&  3.79\% &  78.28\% \\
      & SHSE &  65\% & 55\% &&  96.21\% & 21.72\% \\
  \hline
  \end{tabular}
\end{table}

The most intriguing pattern is observed in the eigenvector $u_2$ of the second largest eigenvalue in 2007 and in the eigenvector $u_3$ of the third largest eigenvalue in 2008. In the eigenvector $u_2$ of 2007, 93.86\% of the positive components correspond to SZSE stocks and 6.14\% to SHSE stocks, while all the negative components are associated with SHSE stocks, as shown in Table~\ref{TB:u2u3}. In the eigenvector $u_3$ of 2008, 3.79\% of the positive components correspond to SZSE stocks and 96.21\% to SHSE stocks, while 78.28\% of the negative components are associated with SZSE stocks. Therefore, the component signs of $u_2$ in 2007 and $u_3$ in 2008 are able to distinguish SZSE stocks from SHSE stocks. This feature is not clearly observed for other eigenvectors. For instance, in the eigenvector $u_2$ of 2008, half of the positive components come from SZSE stocks and 1/3 of the negative components from SZSE stocks.

\section{Market effect}
\label{S1:MarketEffect}

The deviating eigenvalues capture the collective behaviors of different groups of stocks and in particular the largest eigenvalue usually reflects the market effect \cite{Laloux-Cizean-Bouchaud-Potters-1999-PRL}. The characteristic of a market effect is the eigenvector $u_1$ of the largest eigenvalue $\lambda_1$ has roughly equal components on
all of the $N$ stocks, showing a nice linear relationship between the returns of the eigenportfolio constructed from $u_1$ and of the market index \cite{Plerou-Gopikrishnan-Rosenow-Amaral-Guhr-Stanley-2002-PRE}. Usually, other deviating eigenvalues do not reflect a market effect but the comovement of stocks in the same industrial sector \cite{Plerou-Gopikrishnan-Rosenow-Amaral-Guhr-Stanley-2002-PRE}, the same traits shared by stocks \cite{Shen-Zheng-2009a-EPL}, or geographic localization \cite{Dai-Xie-Jiang-Jiang-Zhou-2016-EmpE}. However, it is also possible that other deviating eigenvalues also reflect a market effect, such as the USA housing market \cite{Meng-Xie-Jiang-Podobnik-Zhou-Stanley-2014-SR}.

\begin{figure}[htb]
  \centering
  \includegraphics[width=0.95\linewidth]{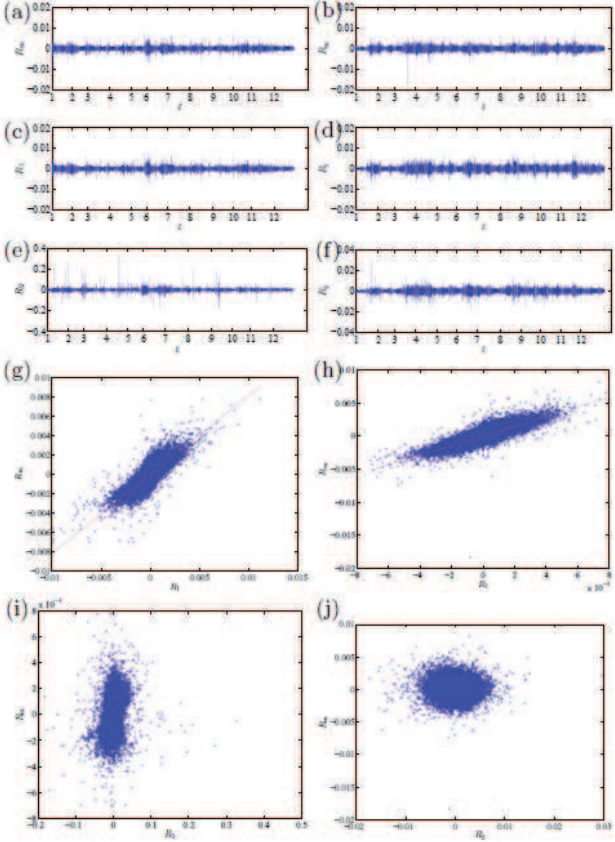}
  \caption{Market effect of the largest eigenvalues $\lambda_1$ of the partial correlation matrices ${\bf{P}}$ in 2007 (left) and 2008 (right). The results for the raw correlation matrices ${\bf{C}}$ are almost the same. The outlier point in (h) and (j) is caused by an abrupt price change at 10:30 on 20 March 2008 shown in (b).} \label{Fig:RMT:Ret:1min:Rm:Rk}
\end{figure}

For each eigenvector ${\mathbf{u}}_k=[u_{k1},\cdots,u_{ki},\cdots, u_{kN}]^{\rm{T}}$ associated with eigenvalue $\lambda_k$, we construct its eigenportfolio, whose returns are calculated by
\begin{equation}
 R_k(t) = \frac{{\mathbf{u}}_k^{\rm{T}}{\mathbf{r}}(t)}{\sum_{i=1}^{N} u_{ki}}
\label{Eq:Partial:Factor}
\end{equation}
where ${\mathbf{r}}(t)=[r_1(t),\cdots,r_i(t),\cdots, r_N(t)]^{\rm{T}}$, and ${\mathbf{u}}_k^{\rm{T}}\cdot{\mathbf{r}}$ denotes the projection of the return time series on the eigenvector ${\mathbf{u}}_{k}$. The return time series $R_m(t)$ of the SSCI, $R_1(t)$ of the first eigenportfolio and $R_2(t)$ of the second eigenportfolio of the partial correlation matrices are plotted in Fig.~\ref{Fig:RMT:Ret:1min:Rm:Rk}(a,c,e) for 2007 and in Fig.~\ref{Fig:RMT:Ret:1min:Rm:Rk}(b,d,f) for 2008. The results are almost the same for the raw correlation matrices, which is natural due to the similarity of corresponding eigenvectors of $\bf{P}$ and ${\bf{C}}$ shown in Fig.~\ref{Fig:RMT:Ret:1min:u1:u5}. It is found that $R_1(t)$ is quite similar to $R_m(t)$ in each year, while $R_2(t)$ is very different from $R_m(t)$.

Fig.~\ref{Fig:RMT:Ret:1min:Rm:Rk}(g-j) present the scatter plots of market returns $R_m(t)$ against the eigenportfolio returns $R_k(t)$ ($k=1,2$) constructed from $\bf{P}$. We find that there is no linear dependence between $R_m(t)$ and $R_2(t)$. No linear dependence is observed either for $R_k(t)$ with $k>2$ (not shown in Fig.~\ref{Fig:RMT:Ret:1min:Rm:Rk}). In contrast, there is a nice linear dependence between $R_m(t)$ and $R_1(t)$ in each year. The slope is 0.815 in 2007 and 0.733 in 2008. For $\bf{C}$, the slope is $0.829\pm0.004$ in 2007 and $0.741\pm0.004$ in 2008. These observations suggest that the largest eigenvalue quantifies a common influence on all stocks, while the rest of deviating eigenvalues contain no information about such a market effect.

The results for the raw correlation matrices $\bf{C}$ are well established for diverse stock markets, especially on the daily level \cite{Plerou-Gopikrishnan-Rosenow-Amaral-Guhr-Stanley-2002-PRE,Shen-Zheng-2009a-EPL}. However, the results for the partial correlation matrices $\bf{P}$ are somewhat surprising. It was expected that the largest eigenvalue is no longer associated with the market, since the effect of the index has been removed \cite{Kenett-Shapira-BenJacob-2009-JPS}. According to our results, this natural conjecture is surprisingly not true. On the contrary, similar phenomena were observed for daily stock returns \cite{Song-2013}.

\section{Eigenvector components and stocks' market capitalizations}
\label{S1:Cap}

We checked the components of eigenvectors associated with the smallest eigenvalues. There are a few components whose magnitudes are significantly greater than the averages. However, we did not observe solid evidence that the correlations of the corresponding return time series of these components are among the largest, which is different from the U.S. stock market \cite{Plerou-Gopikrishnan-Rosenow-Amaral-Guhr-Stanley-2002-PRE} and the global crude oil market \cite{Dai-Xie-Jiang-Jiang-Zhou-2016-EmpE}.

\begin{figure}[htb]
  \centering
  \includegraphics[width=0.48\linewidth]{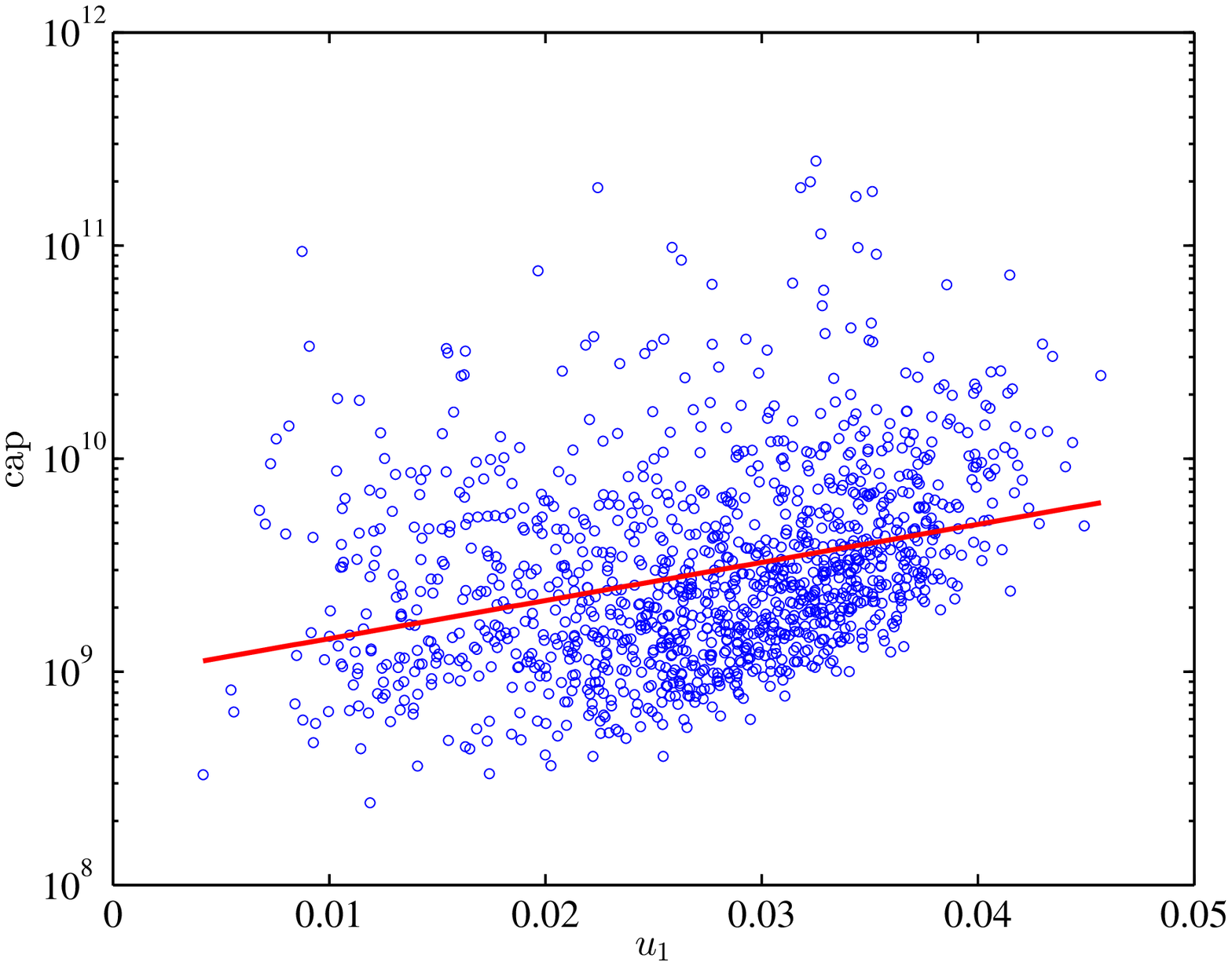}
  \includegraphics[width=0.48\linewidth]{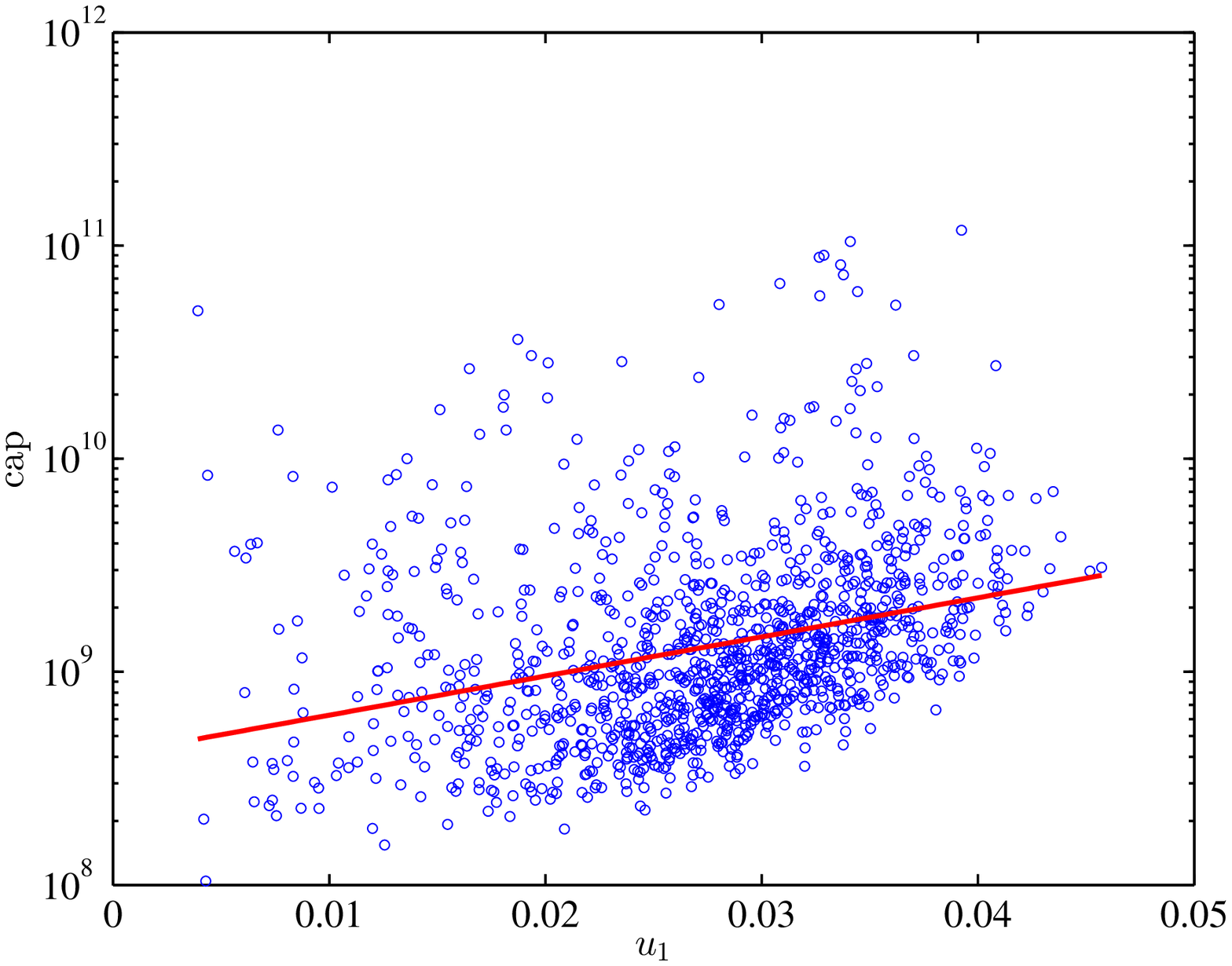}\\
  \includegraphics[width=0.48\linewidth]{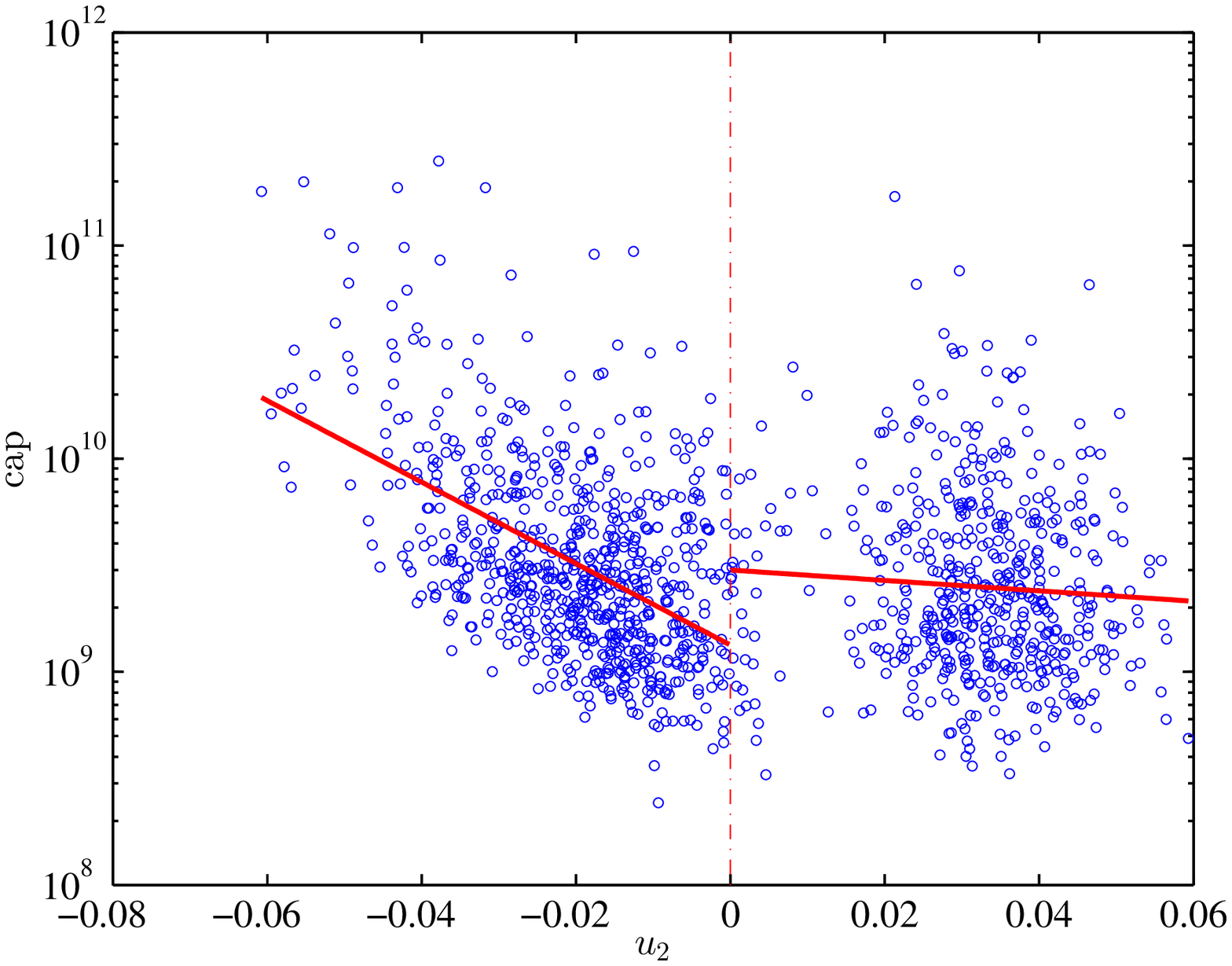}
  \includegraphics[width=0.48\linewidth]{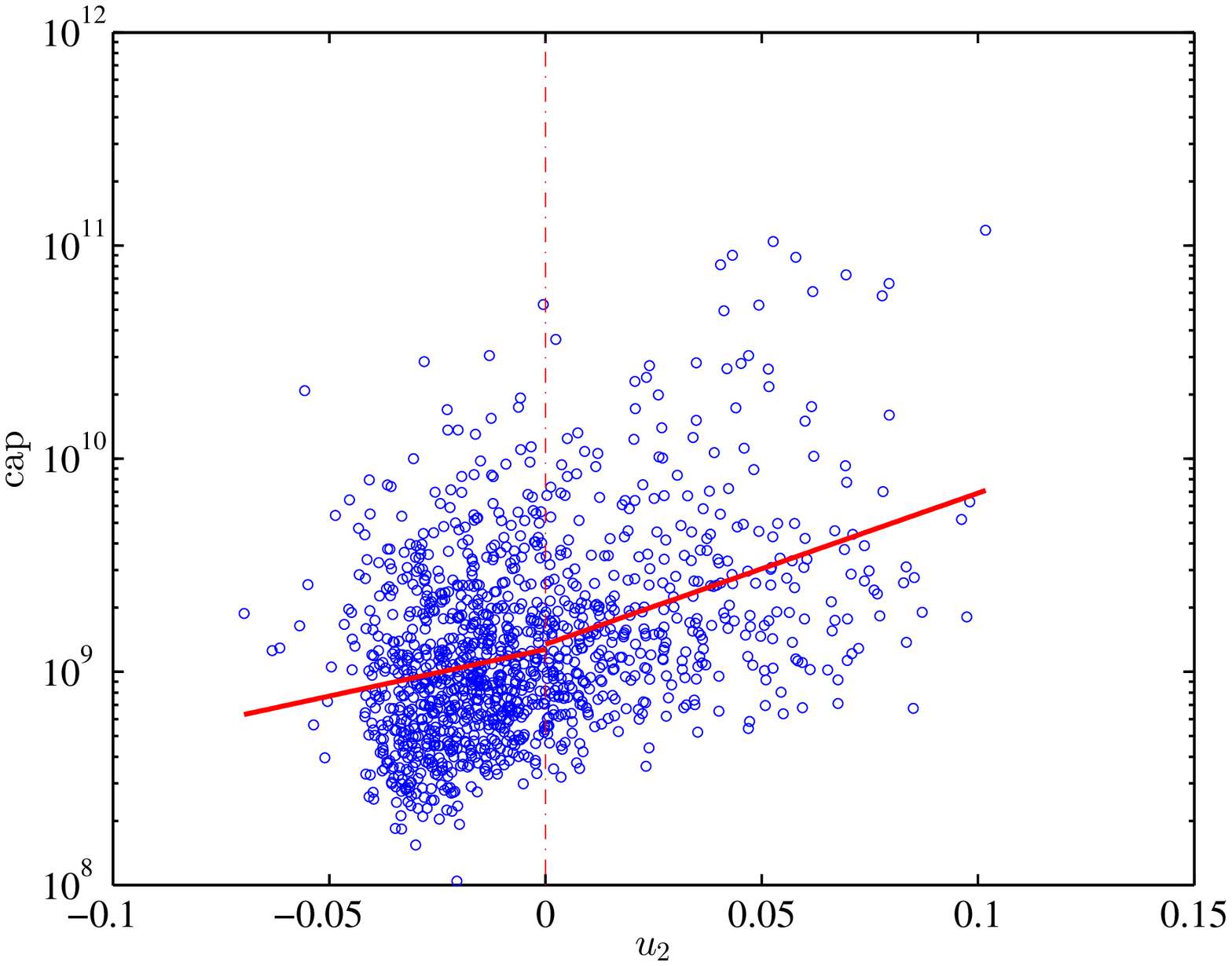}\\
  \includegraphics[width=0.48\linewidth]{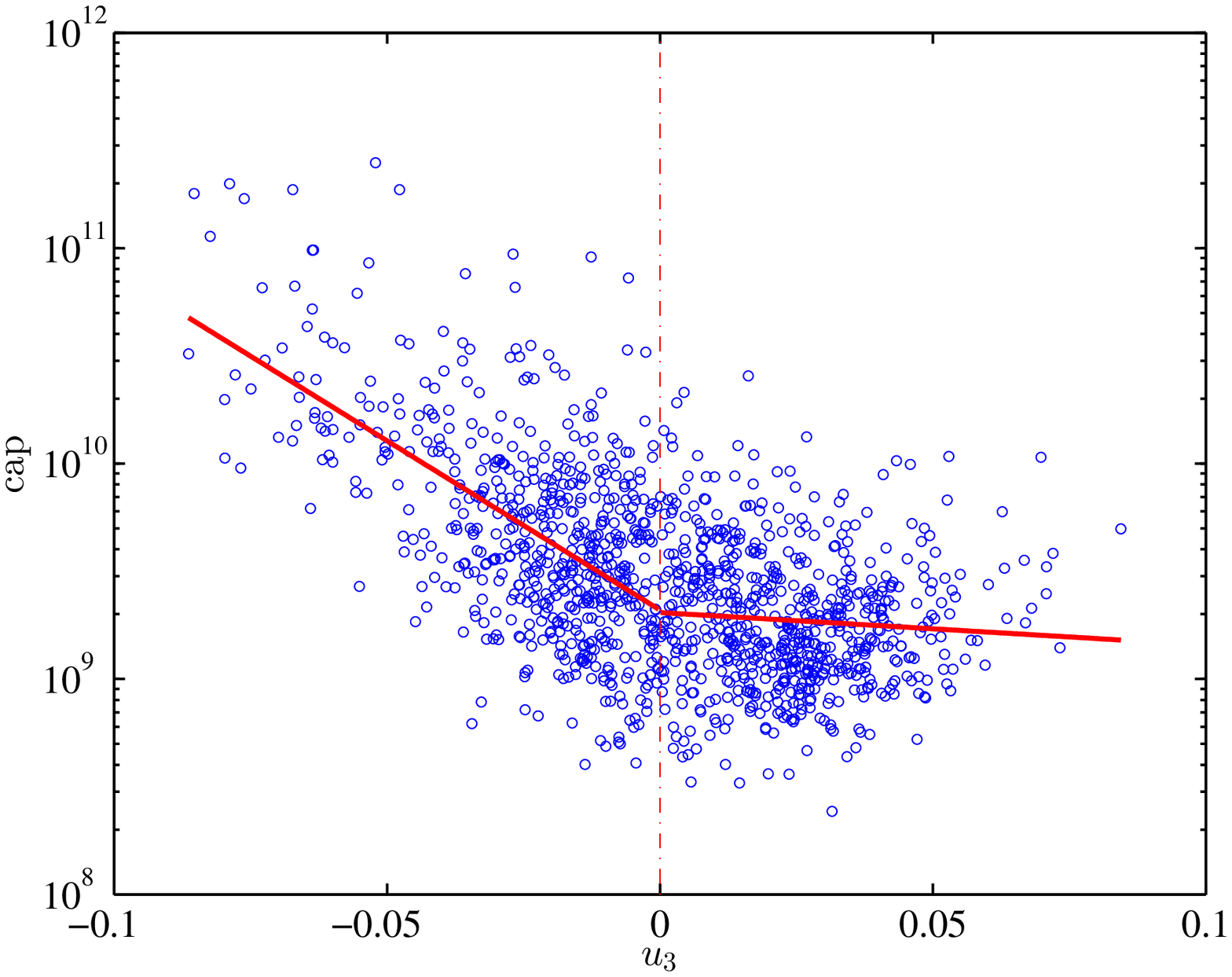}
  \includegraphics[width=0.48\linewidth]{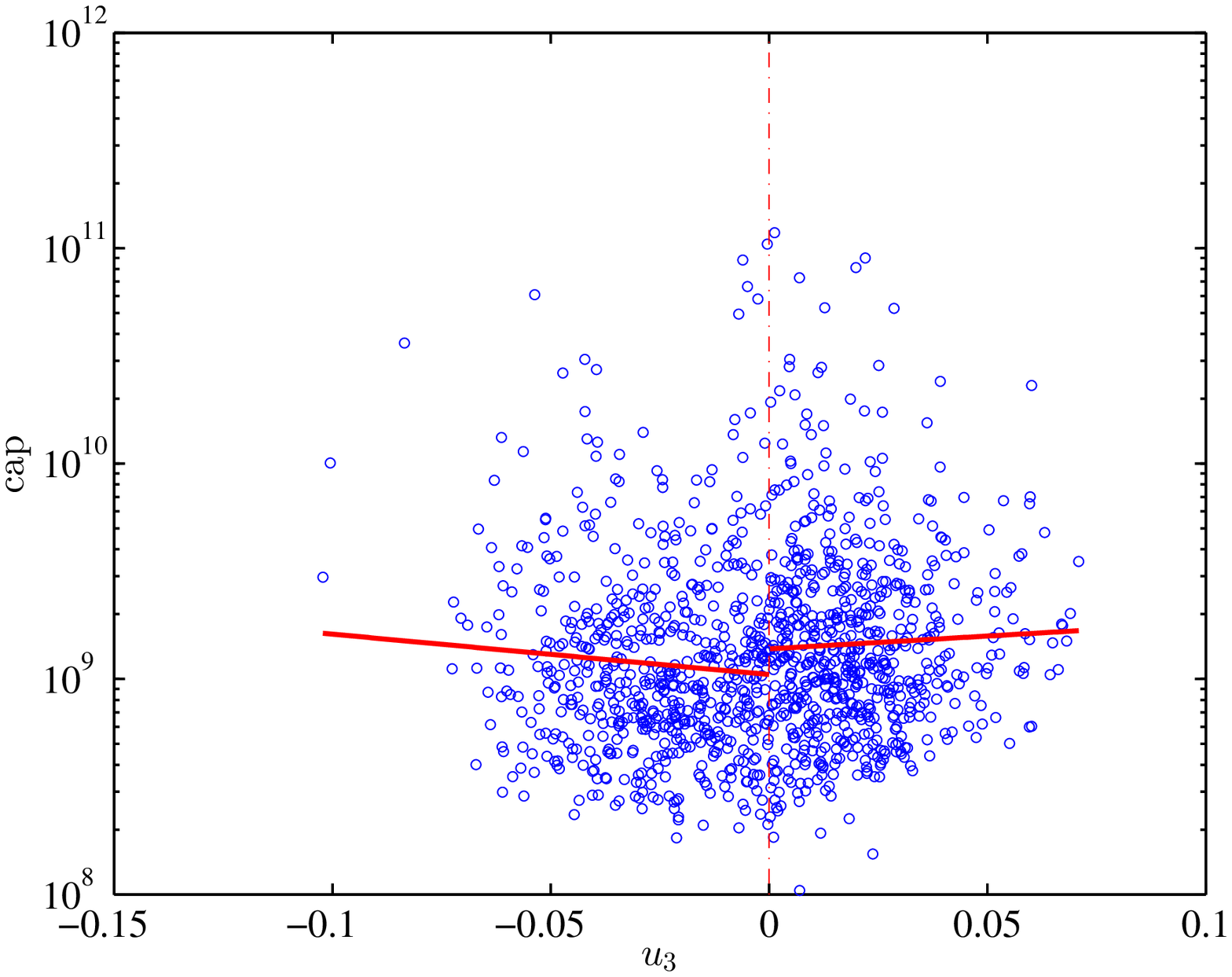}\\
  \vskip  -10.05cm   \hskip   -7.2cm {\small{(a)}}
  \vskip   -0.43cm   \hskip    1.45cm {\small{(b)}}
  \vskip    3.0cm   \hskip   -7.2cm {\small{(c)}}
  \vskip   -0.43cm   \hskip    1.45cm {\small{(d)}}
  \vskip    2.95cm   \hskip   -7.2cm {\small{(e)}}
  \vskip   -0.43cm   \hskip    1.45cm {\small{(f)}}
  \vskip   +2.8cm
  \caption{Relationship between the eigenvector components of the raw correlation matrix and stocks' market capitalizations in 2007 (left) and 2008 (right).} \label{Fig:RMT:Ret:1min:u:Cap}
\end{figure}

We investigated the relationship between the eigenvector components and stocks' market capitalizations. Fig.~\ref{Fig:RMT:Ret:1min:u:Cap} illustrates the results for $u_1$, $u_2$ and $u_3$ of the raw correlation matrix. The results for the partial correlation matrices are similar, as elaborated by the eigenvectors in Fig.~\ref{Fig:RMT:Ret:1min:u1:u5}. The eigenvector components of $u_1$ are positively correlated with the corresponding capitalizations, as shown in Fig.~\ref{Fig:RMT:Ret:1min:u:Cap}(a) and (b). For other largest eigenvalues, such positive correlations between the component magnitudes and capitalizations are observed for positive components or negative components. These findings suggest that stocks with low volatility and high liquidity exhibit high magnitudes of the eigenvector components.

\section{Conclusion}
\label{S1:Conclusion}

We have conducted a comparative analysis of the information contents embedded in the largest eigenvalues of the raw and partial correlation matrices constructed from the 1-min high-frequency returns of Chinese stocks in 2007 and 2008. We identified market correlation structure changes around the Great crash in several aspects. In addition, although the correlation coefficient distributions and the largest eigenvalues of the raw and partial correlation matrix are significantly different in each period, the eigenvectors of the raw and partial correlation matrices in each period are strikingly similar. We found that the largest eigenvalue of each matrix reflects the whole market mode. It is found that the eigenvectors of the second largest eigenvalues in 2007 and of the third largest eigenvalues in 2008 are able to distinguish the stocks from the two exchanges, which are different from the cases of the U.S.A. stock market \cite{Plerou-Gopikrishnan-Rosenow-Amaral-Guhr-Stanley-2002-PRE} and the Chinese stock market when daily returns are analyzed \cite{Shen-Zheng-2009a-EPL,Ren-Zhou-2014-PLoS1}. We also found that the component magnitudes of the some largest eigenvectors are proportional to the stocks' capitalizations.


\acknowledgments

This work was partly supported by the National Natural Science Foundation of China (71532009 and 71131007) and the Fundamental Research Funds for the Central Universities.

\bibliography{E:/Papers/Auxiliary/Bibliography}

\end{document}